\documentclass[numberedappendix,iop]{emulateapj}

\usepackage{graphicx}
\usepackage{amssymb}
\usepackage{epstopdf}
\usepackage{amsmath}
\usepackage{amsfonts}
\usepackage{amssymb}
\usepackage[mathscr]{eucal}
\usepackage{color}
\newcommand{\be}{\begin{equation}}
\newcommand{\ee}{\end{equation}}
\newcommand{\lcdm}{\ensuremath{\Lambda\mathrm{CDM}}}
\newcommand{\fsky}{\ensuremath{f_\mathrm{sky}}}

\newcommand{\nside}{\ensuremath{N_{\rm side}}}
\renewcommand{\ell}{\ensuremath{l}}

\newcommand{\WMAP}{\textsl{WMAP}}
\newcommand{\reffig}[1]{Figure~\ref{#1}}
\newcommand{\reftbl}[1]{Table~\ref{#1}}
\newcommand{\refsec}[1]{\S\ref{#1}}

\shorttitle{WMAP SEVEN-YEAR OBSERVATIONS: POWER SPECTRA AND WMAP-DERIVED PARAMETERS}

\begin{document}

\slugcomment{May 28, 2010}

\title{Seven-Year Wilkinson Microwave Anisotropy Probe 
(\WMAP\altaffilmark{1}) Observations:
Power Spectra and \WMAP-Derived Parameters}
\author{
{{D. Larson}}\altaffilmark{2}, 
{{J. Dunkley}}\altaffilmark{3}, 
{{G. Hinshaw}}\altaffilmark{4}, 
{{E. Komatsu}}\altaffilmark{5}, 
{{M. R. Nolta}}\altaffilmark{6},
{{C. L. Bennett}}\altaffilmark{2}, 
{{B. Gold}}\altaffilmark{2}, 
{{M. Halpern}}\altaffilmark{7}, 
{{R. S. Hill}}\altaffilmark{8}, 
{{N. Jarosik}}\altaffilmark{9}, 
{{A. Kogut}}\altaffilmark{4}, 
{{M. Limon}}\altaffilmark{10}, 
{{S. S. Meyer}}\altaffilmark{11}, 
{{N. Odegard}}\altaffilmark{8}, 
{{L. Page}}\altaffilmark{9}, 
{{K. M. Smith}}\altaffilmark{12}, 
{{D. N. Spergel}}\altaffilmark{12,13},  
{{G. S. Tucker}}\altaffilmark{14}, 
{{J. L. Weiland}}\altaffilmark{8}, 
{{E. Wollack}}\altaffilmark{4}, 
{{E. L. Wright}}\altaffilmark{15}
}

\altaffiltext{1}{\WMAP\ is the result of a partnership between Princeton
University and NASA's Goddard Space Flight Center. Scientific
guidance is provided by the \WMAP\ Science Team.}
\altaffiltext{2}{{Dept. of Physics \& Astronomy, %
                    The Johns Hopkins University, 3400 N. Charles St., %
                    Baltimore, MD  21218-2686}; {dlarson@pha.jhu.edu}}
\altaffiltext{3}{{Astrophysics, University of Oxford, %
                    Keble Road, Oxford, OX1 3RH, UK}}
\altaffiltext{4}{{Code 665, NASA/Goddard Space Flight Center, %
                    Greenbelt, MD 20771}}
\altaffiltext{5}{{Univ. of Texas, Austin, Dept. of Astronomy, %
                    2511 Speedway, RLM 15.306, Austin, TX 78712}}
\altaffiltext{6}{{Canadian Institute for Theoretical Astrophysics, %
                    60 St. George St, University of Toronto, %
                    Toronto, ON  Canada M5S 3H8}}
\altaffiltext{7}{{Dept. of Physics and Astronomy, University of %
                    British Columbia, Vancouver, BC  Canada V6T 1Z1}}
\altaffiltext{8}{{ADNET Systems, Inc., %
                    7515 Mission Dr., Suite A100 Lanham, Maryland 20706}}
\altaffiltext{9}{{Dept. of Physics, Jadwin Hall, %
                    Princeton University, Princeton, NJ 08544-0708}}
\altaffiltext{10}{{Columbia Astrophysics Laboratory, %
                    550 W. 120th St., Mail Code 5247, New York, NY  10027-6902}} 
\altaffiltext{11}{{Depts. of Astrophysics and Physics, KICP and EFI, %
                    University of Chicago, Chicago, IL 60637}}
\altaffiltext{12}{{Dept. of Astrophysical Sciences, %
                    Peyton Hall, Princeton University, Princeton, NJ 08544-1001}}
\altaffiltext{13}{{Princeton Center for Theoretical Physics, %
                    Princeton University, Princeton, NJ 08544}}
\altaffiltext{14}{{Dept. of Physics, Brown University, %
                    182 Hope St., Providence, RI 02912-1843}}
 \altaffiltext{15}{{UCLA Physics \& Astronomy, PO Box 951547, %
                    Los Angeles, CA 90095--1547}}

\begin{abstract}

The \WMAP\ mission has produced sky maps from 7 years of observations at L2.  We
present the angular power spectra derived from the 7-year maps and discuss the
cosmological conclusions that can be inferred from \WMAP\ data alone.  

With the 7-year data, the temperature (TT) spectrum measurement has a
signal-to-noise ratio per multipole that exceeds unity for $\ell<919$; and in
band-powers of width $\Delta l=10$, the signal-to-noise ratio exceeds unity up
to $l=1060$.  The third acoustic peak in the TT spectrum is now well measured by
\WMAP.  In the context of a flat \lcdm\ model, this improvement allows us to
place tighter constraints on the matter density from \WMAP\ data alone,
\ensuremath{\Omega_mh^2 = 0.1334^{+ 0.0056}_{- 0.0055}}, and on the epoch of
matter-radiation equality, \ensuremath{z_{\rm eq} = 3196^{+ 134}_{- 133}}. The
temperature-polarization (TE) spectrum is detected in the 7-year data with a
significance of $20\sigma$, compared to $13\sigma$ with the 5-year data.  We now
detect the second dip in the TE spectrum near $\ell\sim450$ with high
confidence.  The TB and EB spectra remain consistent with zero, thus
demonstrating low systematic errors and foreground residuals in the data.  The
low-$\ell$ EE spectrum, a measure of the optical depth due to reionization, is
detected at $5.5\sigma$ significance when averaged over $\ell=2$--7:
$\ell(\ell+1)C_\ell^{EE}/(2\pi) = 0.074^{+0.034}_{-0.025}$ $\mu$K$^2$ (68\% CL).
We now detect the high-$\ell$, $24 \le \ell \le 800$, EE spectrum at over 8 $\sigma$.
The BB spectrum, an important probe of gravitational waves from inflation,
remains consistent with zero; when averaged over $\ell=2$--7,
$\ell(\ell+1)C_\ell^{BB}/(2\pi) < 0.055$ $\mu$K$^2$ (95\% CL). The upper limit
on tensor modes from polarization data alone is a factor of 2 lower with the
7-year data than it was using the 5-year data \citep{komatsu/etal:prep}.

The data remain consistent with the simple \lcdm\ model: the best-fit TT
spectrum has an effective $\chi^2$ of 1227 for 1170 degrees of freedom, with a
probability to exceed  of 9.6\%.  The allowable volume in the 6-dimensional
space of \lcdm\ parameters has been reduced by a factor of 1.5 relative to the
5-year volume, while the \lcdm\ model that allows for tensor modes and a running
scalar spectral index has a factor of 3 lower volume when fit to the 7-year
data.  We test the parameter recovery process for bias and find that the scalar
spectral index, \ensuremath{n_s}, is biased {\it high}, but only by 0.09 $\sigma$,
while the remaining parameters are biased by $<0.15$ $\sigma$.

The improvement in the third peak measurement leads to tighter lower limits from
\WMAP\ on the number of relativistic degrees of freedom (e.g., neutrinos) in the
early universe: \ensuremath{N_{\rm eff} > 2.7\ \mbox{(95\% CL)}}.  Also,
using \WMAP\ data alone, the primordial helium mass fraction is found to be
\ensuremath{Y_{\rm He} = 0.28^{+ 0.14}_{- 0.15}}, and with data from
higher-resolution CMB experiments included, we now establish the existence of
pre-stellar helium at $>3\sigma$ \citep{komatsu/etal:prep}.  These new \WMAP\
measurements provide important tests of Big Bang cosmology. 

\end{abstract}

\section{INTRODUCTION}

The Wilkinson Microwave Anisotropy Probe
\citep[\WMAP][]{bennett/etal:2003,bennett/etal:2003b} is a Medium-Class Explorer
(MIDEX) satellite aimed at understanding cosmology through full-sky observations
of the cosmic microwave background (CMB).  The \WMAP\ full-sky maps of the
temperature and polarization anisotropy in five frequency bands provide our most
accurate view to date of conditions in the early universe.  The \WMAP\
instrument is composed of 10 differencing assemblies (DAs) spanning 5
frequencies from 23 to 94 GHz \citep{bennett/etal:2003}: one DA each at 23 GHz
(K1) and 33 GHz (Ka1), two each at 41 GHz (Q1,Q2) and 61 GHz (V1,V2), and four
at 94 GHz (W1--W4).   Each DA is formed from two differential radiometers which
are sensitive to  orthogonal linear polarization modes; \WMAP\ measures both
temperature and polarization at each frequency. The multi-frequency data
facilitate the separation of the CMB signal from foreground emission arising
both from our Galaxy and from extragalactic sources.  The CMB angular power
spectrum derived from these maps exhibits a highly coherent acoustic peak
structure which makes it possible to extract a wealth of information about the
composition and history of the universe, as well as the processes that seeded
the fluctuations.

With accurate measurements of the first few peaks in the angular power spectrum,
CMB data have enabled the following advances in our understanding of cosmology
\citep{spergel/etal:2003,spergel/etal:2007,dunkley/etal:2009,komatsu/etal:2009}:
the dark matter must be non-baryonic and interact only weakly with atoms and
radiation; the density of atoms in the universe is known to 3\%, and accords
well with Big Bang Nucleosynthesis; the measured acoustic scale at $z=1090$,
combined with the local distance scale and Baryon Acoustic Oscillation (BAO) data, 
demonstrates that the universe is
spatially flat, to within 1\%; the Hubble constant is determined to 3\% using only acoustic
fluctuation data (CMB+BAO), and it accords well with local measurements; the
primordial fluctuations are adiabatic and Gaussian, and the spectrum is slightly tilted.

The statistical properties of the CMB fluctuations measured by \WMAP\ are close
to Gaussian with random phase 
\citep{komatsu/etal:2003,spergel/etal:2007,komatsu/etal:2009}.  
There are several hints of possible deviations from
this case as discussed in \citet{bennett/etal:prep,
komatsu/etal:prep}.  If the fluctuations are Gaussian and random phase, then
their statistical information content is completely determined by the angular
power spectra of the sky maps.

This paper derives the angular power spectra from the \WMAP\ 7-year sky maps and
presents the cosmological parameters that can be determined from them.  The new
results improve upon previous results in many ways: additional data reduces the
random noise, which is especially important for studying the temperature signal
on small angular scales and the polarization signal on large angular scales;  W
band data is now incorporated in the TE spectrum measurement to improve
precision; new simulations have been carried out to test the accuracy of
parameter recovery and to test a model's goodness of fit.  The result is the
most accurate full sky measurement to date of CMB anisotropy down to an angular scale of
$\sim0.25^{\circ}$.

This paper is one of six that accompany the 7-year \WMAP\ data release.
\citet{jarosik/etal:prep} discuss the 7-year map making process, systematic
error limits, and basic results. \citet{gold/etal:prep} discuss galactic
foreground emission, and its removal in 7-year data. \citet{bennett/etal:prep}
discuss possible anomalies in the \WMAP\ CMB maps.  \citet{komatsu/etal:prep}
discuss the interpretation of the \WMAP\ data, in combination with other
relevant cosmological data. \citet{weiland/etal:prep} discuss the \WMAP\
measurements of the outer planets and selected bright sources for use as
microwave calibrators.

The layout of this paper is as follows. In \S\ref{sec_power_spectra}, we present
the \WMAP\ 7-year power spectra.  In \S\ref{sec_parameter_recovery}, we discuss
simulations that were performed to test for bias in our cosmological parameter
fits. In \S\ref{sec_parameters} we discuss cosmological conclusions that can be
drawn from \WMAP\ data alone, and in \S\ref{sec_goodness_of_fit} we discuss the
goodness of fit of the 6-parameter \lcdm\ theory.  We conclude in \S\ref{sec_conclusion}.

\section{SEVEN-YEAR POWER SPECTRA}
\label{sec_power_spectra}

In this section we present the temperature and polarization power spectra
derived from the 7-year sky maps and compare them to the 5-year spectra.

\subsection{Definitions and Methodology}

Since \WMAP\ measures both temperature and polarization, there are multiple
power spectra to consider.  On a sphere, the temperature field can be decomposed
into spherical harmonics,
\be
T(\hat{n}) = \sum_{\ell=0}^\infty \sum_{m=-\ell}^{\ell} a_{T,\ell m} Y_{\ell m}(\hat{n}),
\ee
where $\hat{n}$ is a unit direction vector and $a_T$ refers specifically to the
temperature field. Likewise, the Q and U Stokes parameters for linear
polarization can be decomposed into complex spin-2 harmonics
\citep{newman/penrose:1966, goldberg/etal:1967},
\be
Q(\hat{n}) + i U(\hat{n}) = \sum_{\ell=2}^\infty \sum_{m=-\ell}^{\ell} a_{2,\ell m} \,_2Y_{\ell m}(\hat{n}).
\ee
The spin-2 coefficients can then be combined to represent polarization modes
that have no curl (E modes) and modes that have no divergence (B modes).
These are given by the coefficients \citep{zaldarriaga/seljak:1997, larson:2006},
\begin{eqnarray}
a_{E,\ell m} & = & -\frac{a_{2,\ell m} + (-1)^m a^*_{2,\ell\; -m}}{2} \\
a_{B,\ell m} & = & -\frac{a_{2,\ell m} - (-1)^m a^*_{2,\ell\; -m}}{2i}
\end{eqnarray}
(\citet{kamionkowski/kosowsky/stebbins:1997} use an alternative approach.)
The angular power spectra are related to these modes according to
\be
C^{XY}_\ell = \frac{1}{2\ell+1}\sum_{m=-\ell}^\ell a_{X,\ell m} a^*_{Y,\ell m}
\ee
where X, Y = T, E, or B.  
The data
are currently consistent with being isotropic and Gaussian distributed, but this condition
should continue to be tested \citep{bennett/etal:prep}.

There are 6 independent power spectra than can be constructed from the
temperature and polarization data, TT, TE, TB, EE, EB, and BB, though in
theories in which parity is conserved, TB and EB are expected to be zero.  In general,
foreground signals (and systematic effects) can produce non-zero TB, EB, and BB
so these spectra provide a good test for residual polarization contamination.

For the 7-year analysis, we use the same combination of estimators that were
used with the 5-year data \citep{nolta/etal:2009}.  This combination is a
trade-off between statistical accuracy and computational speed.  We present new
tests of the accuracy of the likelihood function constructed from these
estimators in \S\ref{sec_parameter_recovery}.  To summarize the combination:
for low-$\ell$ TT ($ \le 32$) we compute the likelihood of a model directly
from the 7-year Internal Linear
Combination (ILC) maps \citep{gold/etal:prep}.  For high-$\ell$ ($\ell > 32$) TT we use the
MASTER pseudo-$C_\ell$ quadratic estimator \citep{hivon/etal:2002}.  For
low-$\ell$ polarization, $\ell \le 23$ TE, EE, \& BB, we use the pixel-space
estimator described in \citet{page/etal:2007}, and for high-$\ell$ TE ($\ell >
23$) we use the MASTER quadratic estimator.

\subsection{Changes Affecting the 7-Year Spectra}

Several data processing and analysis changes were applied to the 7-year data
which resulted in improvements beyond those which would be expected from
additional integration time.

\subsubsection{Map-Making with Asymmetric Masking}

A new map-making technique was adopted for the 7-year data which combines optimal
noise handling with ``asymmetric'' data masking \citep{jarosik/etal:prep}.  With
this change, certain regions in the 7-year maps employ more
data samples than they would have with the previous pipeline.  These ``Galactic
echo'' regions are thus more sensitive than a simple 5-year to 7-year integration
time scaling would predict.

\subsubsection{Multipole Range}

The additional sensitivity afforded by more data has made it possible to extend
the usable multipole range in the power spectra and likelihood code.  For the TT
data we extend the upper multipole limit, $\ell_{\rm max}$, from 1000 to 1200. 
For the TE spectrum, we have determined that high-$\ell$ W-band polarization
data are sufficiently free from systematic effects that they can be employed in
the TE spectrum estimate \citep{jarosik/etal:prep}.  This significantly improves
the sensitivity in the 7-year TE spectrum and allows us to extend the TE
multipole limit from 450 to 800.

\subsubsection{Mask and $f_{\rm sky}$}
\label{sec_mask_fsky}

The 7-year sky masks have been augmented slightly using a $\chi^2$ analysis of
the Q--V \& V--W difference maps, {\it after} the normal template cleaning had
been applied \citep{jarosik/etal:prep, gold/etal:prep}.  This results in a
slightly more conservative mask which decreases the unmasked sky fraction by
$\sim3$\% (from 81.7\% to 78.3\% for the KQ85 cut---the new cut is denoted
KQ85y7).  Given the $\chi^2$ threshold applied during the construction of the
extended mask, residual foreground signals outside the mask are essentially
undetectable on the scale of the instrument noise in a $\sim$2$^{\circ}$ pixel,
so the data are fractionally more robust to foreground contamination.

The power spectrum sensitivity depends on sky cut according to $\Delta C_\ell
\propto f_{\rm sky}^{-1}$ where $f_{\rm sky}$ is (approximately) the fraction of
sky that survives the cut \citep{hinshaw/etal:2003, verde/etal:2003}.  In
practice, $f_{\rm sky}$ is a function of $\ell$ that is calibrated with
simulations, so there is a different constant of proportionality at each $\ell$,
but it scales with the fraction of usable sky area.  Thus the increased sky mask
results in a slight loss of sensitivity in the TT spectrum for $\ell \lesssim
550$, where the spectrum is sky variance limited.

For the TE spectrum, we have generated new simulations to test the calibration
of $f_{\rm sky,TE}$ which enters into the error propagation from sky maps to
spectra.  The mean $\chi^2$ deduced from our simulations was 760 for
spectra with 777 degrees of freedom, a factor of 1.022 too low, indicating that
our previous TE error estimate was a factor of 1.011 too high.  Therefore we
have scaled $f_{\rm sky,TE}$ by 1.011 to produce a unit mean $\chi^2$ per degree of
freedom, following the precedent used to calibrate $f_{\rm sky,TT}$ from
simulations \citep{verde/etal:2003}.  In \S\ref{sec_goodness_of_fit} we
present the TE $\chi^2$ of the 7-year flight data and conclude that the \lcdm\
model fits the TE data well.

\subsection{Temperature (TT) Spectrum}
\label{sec_tt_spec}

\begin{figure*}
\epsscale{1.0}
\plotone{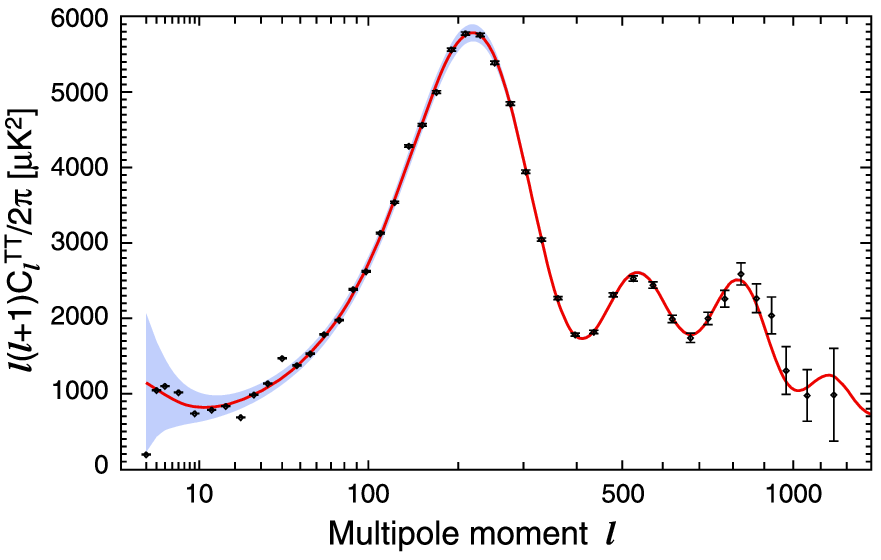}
\caption{\label{figure_tt}
The 7-year temperature (TT) power spectrum from \WMAP. The third acoustic peak
and the onset of the Silk damping tail are now well measured by \WMAP.  The curve is the
\lcdm\ model best fit to the 7-year \WMAP\ data:
\ensuremath{\Omega_bh^2}$=0.02270$,
\ensuremath{\Omega_ch^2}$=0.1107$, 
\ensuremath{\Omega_\Lambda}$=0.738$,
\ensuremath{\tau}$=0.086$, 
\ensuremath{n_s}$=0.969$,
\ensuremath{\Delta_{\cal R}^2}$=2.38\times10^{-9}$, and
\ensuremath{A_{\rm SZ}}$=0.52$.
The plotted errors include instrument noise, but not the small, correlated
contribution due to beam and point source subtraction uncertainty.  The gray
band represents cosmic variance. A complete error treatment is incorporated in
the \WMAP\ likelihood code.  The points are binned in progressively larger
multipole bins with increasing $\ell$; the bin ranges are included in the 7-year
data release.}
\end{figure*}

\begin{figure}
\epsscale{1.2}
\plotone{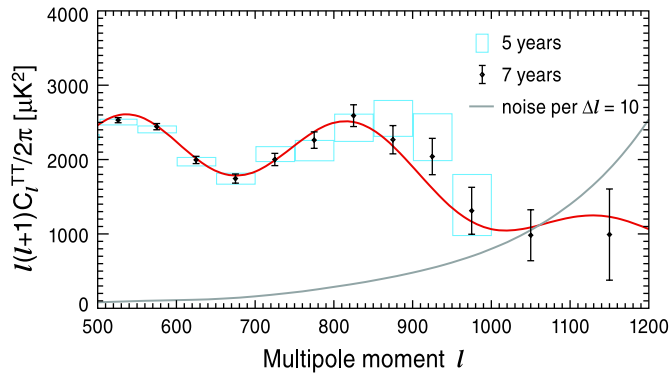}
\caption{\label{fig_spec_tt_highl}
The high-$\ell$ TT spectrum measured by \WMAP, showing the improvement with 7
years of data.  The points with errors use the full data set while the boxes
show the 5-year results with the same binning.  The TT measurement is improved
by $>$30\% in the vicinity of the third acoustic peak (at $\ell \approx 800$), while the 2 bins from
$\ell$ = 1000--1200 are new with the 7-year data analysis.  
}
\end{figure}

For $\ell \le 32$, the spectrum is obtained using a Blackwell-Rao estimator
applied to a chain of Gibbs samples \citep{wandelt/larson/lakshminarayanan:2004,
jewell/levin/anderson:2004, chu/etal:2005,  dunkley/etal:2009}  based on the
7-year ILC map and the KQ85y7 mask.  The specifications used to sample the map
are described in \citet{dunkley/etal:2009}.  For $\ell>32$, the spectrum was
derived from the MASTER pseudo-$C_\ell$ quadratic estimator applied to the
7-year, template-cleaned V and W-band maps \citep{gold/etal:prep}. (The MASTER
spectrum is technically derived from $\ell=2$ to 1200, then the $\ell=2-32$
portion is discarded, but correlations induced by mode coupling are retained for
$\ell>32$.)  The pseudo-$a_{\ell m}$ coefficients were computed from the $\nside
= 1024$ maps for each single year and each single-DA, V1-W4.  For $\ell<600$ the
coefficients were evaluated with uniform pixel weights, while inverse-noise
weights were used for $\ell>600$. (The transition was made at $\ell=500$ in the
5-year analysis.)  As noted above, we adopt a slightly larger sky mask, denoted
KQ85y7.  

The pseudo-$C_\ell$ cross-power spectra are computed from all off-diagonal
pairs of pseudo-$a_{\ell m}$ coefficients,
\be
{\tilde C}^{(ij)}_\ell = \frac{1}{2\ell+1} 
\sum_{m=-\ell}^{\ell}  {\tilde a}^{(i)}_{\ell m} {\tilde a}^{*(j)}_{\ell m}
\ee
where $i$, $j$ refer to a DA-year combination \citep{hinshaw/etal:2007} and the
tilde indicates a pseudo-quantity (spectrum or coefficient).  These component
pseudo-spectra are deconvolved using the MASTER formalism\footnote{
In principle, we can obtain a modestly more sensitive spectrum estimate at intermediate 
multipoles by employing $C^{-1}$ weighting in the computation of the pseudo-$a_{\ell m}$.  We 
are currently developing such code for use in cross-power spectra with the intention of applying 
it to the final 9-year data.}, and the results are
combined by band into VV, VW, and WW spectra for the purposes of removing the
residual point source amplitude. The unresolved point source contribution to the
sky continues to be treated as a power law in thermodynamic temperature, falling
as  $\nu^{-2.09}$ \citep{nolta/etal:2009}, but see \citet{colombo/pierpaoli:2010} for
an alternative approach to the spectral dependence. 
Using the same fitting methodology as
in the 5-year analysis, we find its amplitude to be $10^3 A_{\rm ps}=9.0\pm
0.7\,\mu{\rm K}^2\;{\rm sr}$, when fit to the 7-year Q, V, and W band spectra
evaluated with the KQ85y7 mask.  (Most of the cosmological parameters reported in this paper 
were fit using a preliminary version of the likelihood that had a small masking error that 
produced a slightly biased TT spectrum at high-$\ell$ and a correspondingly higher residual source 
amplitude, which mostly compensated for the bias. We have checked that substituting the 
correct TT spectrum has a negligible effect on the parameter fits.)  After this source
model is subtracted from each band, the spectra are combined to form our best
estimate of the CMB signal, shown in Figure~\ref{figure_tt}.

The 7-year power spectrum is cosmic variance limited, i.e., cosmic variance
exceeds the instrument noise, up to $\ell=548$. (This limit is slightly model
dependent and can vary by a few multipoles.)  The spectrum has a signal-to-noise
ratio greater than one per $\ell$-mode up to $\ell=919$, and in band-powers of
width $\Delta l=10$, the signal-to-noise ratio exceeds unity up to $l=1060$. The
largest improvement in the 7-year spectrum occurs at multipoles $\ell>600$ where
the uncertainty is still dominated by instrument noise.  The instrument noise
level in the 7-year spectrum is 39\% smaller than with the 5-year data, which
makes it worthwhile to extend the \WMAP\ spectrum estimate up to $\ell=1200$ for
the first time.  
See \reffig{fig_spec_tt_highl} for a comparison of the 7-year error bars
to the 5-year error bars.
The third acoustic peak is now well measured and the onset of
the Silk damping tail is also clearly seen by \WMAP.  As we show in
\S\ref{sec_parameters}, this leads to a better measurement of
\ensuremath{\Omega_mh^2} and the epoch of matter-radiation equality,
\ensuremath{z_{\rm eq}}, which, in turn, leads to better constraints on the effective
number of relativistic species, \ensuremath{N_{\rm eff}}, and on the primordial helium
abundance, \ensuremath{Y_{\rm He}}.  The improved sensitivity at high $\ell$ is also
important for higher-resolution CMB experiments that use \WMAP\ as a primary
calibration source.

\subsection{Temperature-Polarization (TE, TB) Cross Spectra}
\label{sec_tp_spec}

\begin{figure}
\epsscale{1.2}
\plotone{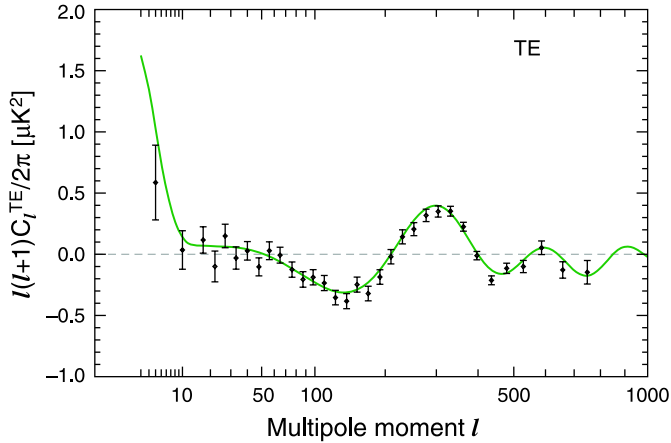}
\caption{\label{figure_te}
The 7-year temperature-polarization (TE) cross-power spectrum measured by
\WMAP.  The second trough (TE$<$0) in the spectrum in the vicinity of $l=450$ is
now clearly detected. The green curve is the \lcdm\ model best fit to the 7-year
\WMAP\ data, as in Figure~\ref{figure_tt}. The plotted errors depict the
diagonal elements of the covariance matrix and include both cosmic variance and
instrument noise.  A complete error treatment is incorporated in the \WMAP\
likelihood code.  Note that the plotted spectrum is $(\ell+1)C^{\rm TE}_\ell/(2\pi)$,
and not $\ell(\ell+1)C^{\rm TE}_\ell/(2\pi)$.}
\end{figure}

\begin{figure*}
\epsscale{0.8}
\plotone{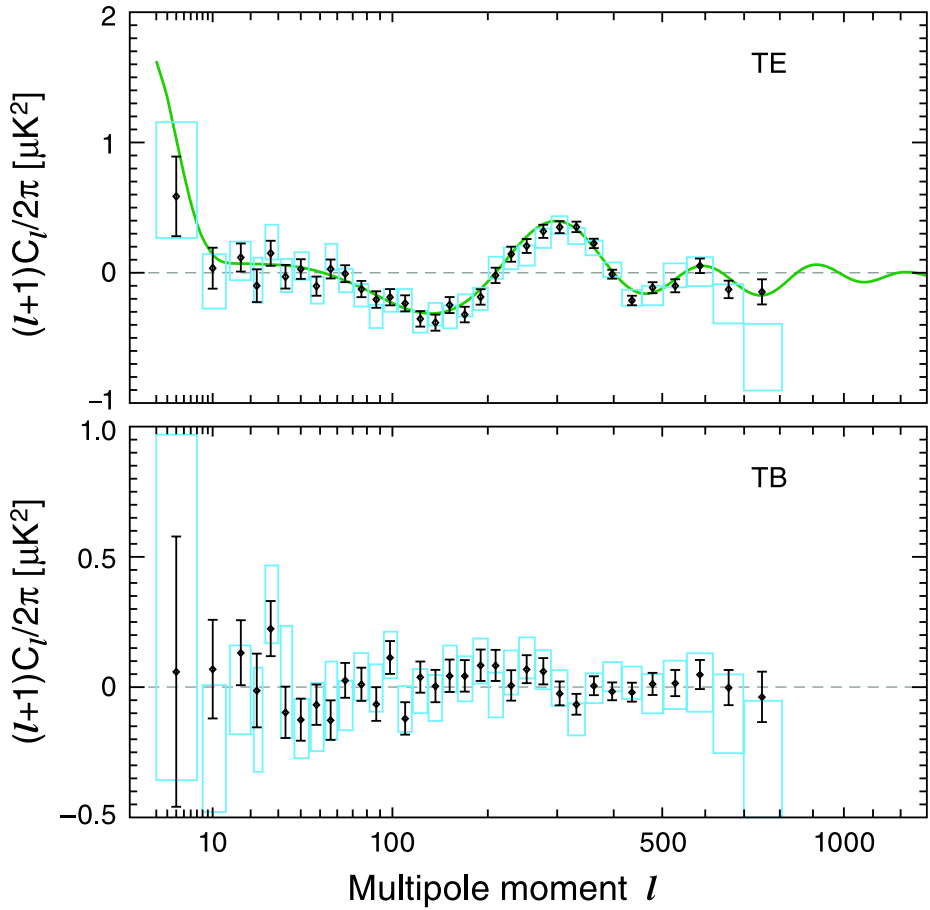}
\caption{\label{fig_spec_57}
The TE and TB high-$\ell$ spectra measured by \WMAP, showing the improvement with 7
years of data.  The points with errors use the full data set while the boxes
show the 5-year results with the same binning.  The spectra
are greatly improved by the addition of W-band data.  
The non-detection of
TB signal is expected; it provides a good check of systematic errors and
foreground residuals, and can be also used to set limits on polarization
rotation due to parity-violating effects (\S\ref{sec_tp_spec} and
\citet{komatsu/etal:prep}).}
\end{figure*}

\begin{figure}
\epsscale{1.2}
\plotone{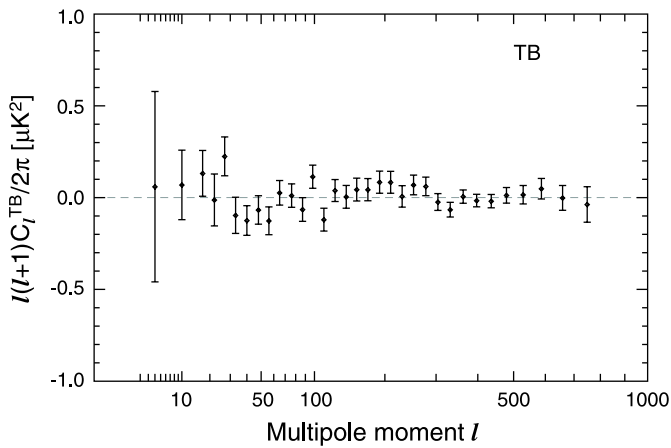}
\caption{\label{figure_tb}
The 7-year temperature-polarization (TB) cross-power spectrum measured by
\WMAP.  This spectrum is predicted to be zero in the basic \lcdm\ model and the
measured spectrum is consistent with zero.  TB provides a useful null test for
systematic errors and foreground residuals.  \citet{komatsu/etal:prep} use the
TB and TE spectra to place an upper limit on polarization rotation due to
parity-violating effects. 
The TB $\chi^2$ for the
null hypothesis (TB=0) is 793.5 for 777 degrees of freedom.  The probability to
exceed that amount is 33\%.
Note that the plotted spectrum is $(\ell+1)C^{\rm
TB}_\ell/(2\pi)$, and not $\ell(\ell+1)C^{\rm TB}_\ell/(2\pi)$.}
\end{figure}

The 7-year temperature-polarization cross power spectra were formed using the
same methodology as the 5-year spectrum \citep{page/etal:2007,nolta/etal:2009}. 
For $\ell \le 23$ the cosmological model likelihood is estimated directly from
low-resolution temperature and polarization maps. The temperature input is a
template-cleaned, co-added V+W band map, while the polarization input is a
template-cleaned, co-added Ka+Q+V band map \citep{gold/etal:2009}. In this
regime, the spectrum can be inferred from the conditional likelihood of $C_l$
values (individual or binned), but these estimates are only used for
visualization.

For $\ell>23$, the temperature-polarization spectra are derived using the MASTER
quadratic estimator, extended to include polarization data
\citep{page/etal:2007}.  (As above, the MASTER spectrum is evaluated from
$\ell=2$, but the result from $\ell=2-23$ is discarded.) The temperature input
is a template-cleaned, co-added V+W band map, while the polarization input is a
template-cleaned, co-added Q+V+W band map.  The inclusion of W-band data in
the high-$\ell$ TE and TB spectra is new with the 7-year data release
\citep{jarosik/etal:prep}.  Since the W band radiometers have the highest
angular resolution, the inclusion of W band significantly enhances the
sensitivity of these high-$\ell$ spectra.

The 7-year TE spectrum measured by \WMAP\ is shown in Figure~\ref{figure_te}.  
For all except the first bin, the MASTER values and their Gaussian errors are
plotted. The first bin shows the conditional maximum likelihood value 
based on the pixel likelihood mentioned above. 
The slight adjustment for $f_{\rm sky,TE}$ is included in the error bars.
With two
additional years of integration and the inclusion of W band data, we now detect
the TE signal with a significance of 20$\sigma$, up from 13$\sigma$ with the
5-year data.  Indeed, for $10<\ell<300$, the TE error is less than 65\% of the 5-year
value, and for $\ell>300$ the sensitivity improvement is even larger due to W
band's finer resolution.  At $\ell=800$ the 7-year TE error is 36\% of the
5-year value. A qualitatively new feature seen in the 7-year spectrum is a
second trough (TE$<$0) near $\ell=450$.  
See \reffig{fig_spec_57} for a comparison of the 7-year to 5-year error bars,
for the TE and TB spectra.
Overall, the TE data are quite
consistent with the simplest 6-parameter \lcdm\ model; we discuss its
goodness-of-fit in \S\ref{sec_goodness_of_fit}.

The observed TE signal is the result of a specific polarization pattern around hot
and cold spots in the temperature anisotropy.  In particular, the acoustic peak
structure in TE corresponds to a series of concentric rings of alternating radial
and tangential polarization (relative to a radial reference direction). 
\citet{komatsu/etal:prep} perform a stacking analysis of the 7-year temperature and
polarization maps and show that the effect is detected in the 7-year \WMAP\ sky
maps with a significance of 8$\sigma$.  

The 7-year TB spectrum measured by \WMAP\ is shown in Figure~\ref{figure_tb}. 
In this case, because the signal-to-noise ratio is low, the MASTER points and
their Gaussian errors are plotted over the full $\ell$ range, including the first
bin.  The measured spectrum is consistent with zero: the $\chi^2$ for the
null hypothesis (TB=0) is 793.5 for 777 degrees of freedom.  The probability to
exceed that amount is 33\%.  The
absence of a detectable signal is consistent with the \lcdm\ model, which
predicts zero.  It is also an indication that systematic errors and foreground
contamination are not significant at the level of $\sim 0.1$ $\mu$K$^2$ in
$(\ell+1)C_\ell^{\rm TB}$.

\citet{komatsu/etal:prep} use the 7-year TE and TB data to place limits on
polarization rotation due to parity violating effects.  Polarization rotation
would cause TE signal generated at the last scattering surface to transform into
observed TB power.  The absence of TB signal leads to an
upper limit on rotation of $\Delta\alpha = -1.1^{\circ} \pm 1.4^{\circ} ({\rm
stat}) \pm 1.5^{\circ} ({\rm sys})$.

\subsection{Polarization (EE, EB, BB) Spectra}
\label{sec_pp_spec}

\begin{figure*}
\epsscale{1.0}
\plotone{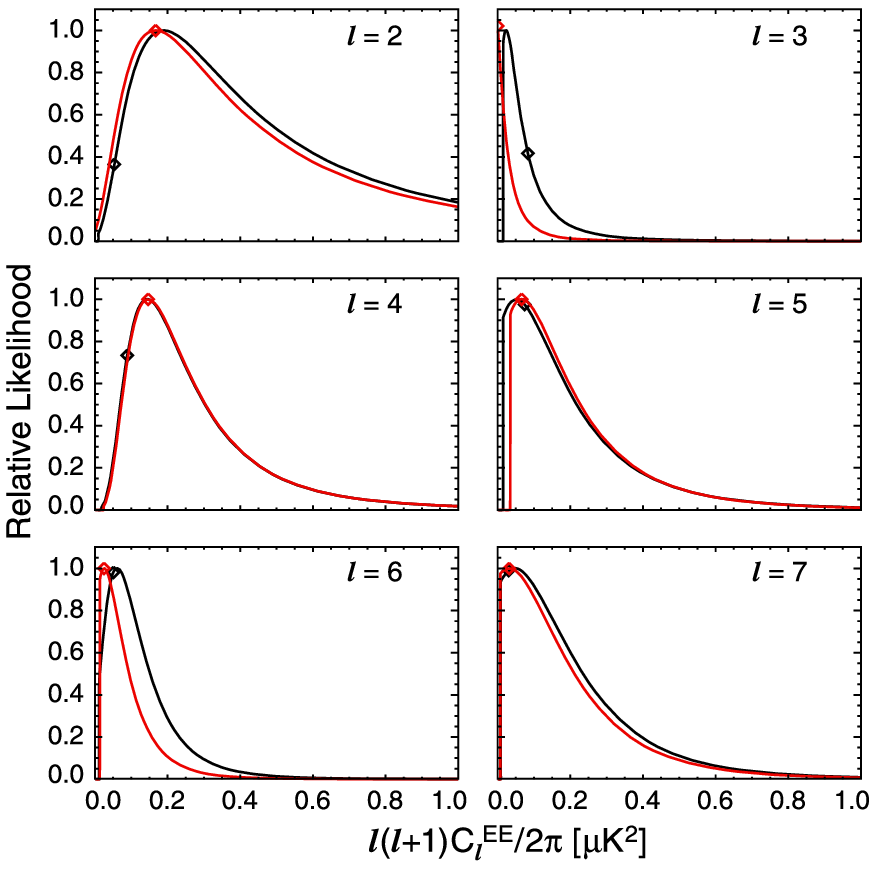}
\caption{\label{figure_low_EE}
Conditional likelihoods of $\ell(\ell+1)C_\ell^{\rm EE}/(2\pi)$ for $\ell$ = 2--7,
computed with the \WMAP\ likelihood code. The Ka, Q, and V bands contribute
to the low-$\ell$ polarized pixel likelihood.
In each panel, the black curve is the
conditional likelihood for a given multipole when all other multipoles are held
fixed at the value of the best-fit \lcdm\ model (indicated by the black
diamonds). The red curve in each panel is the conditional likelihood when all
other multipoles are held fixed at the maximum likelihood value of the spectrum,
indicated by the red diamonds.  The maximum likelihood spectra were determined
by a numerical  maximization of the \WMAP\ likelihood code, for $\ell$ = 2--10,
for TT, TE, EE, and BB.  Points with  $\ell>10$ were fixed at the best-fit
\lcdm\ value.}
\end{figure*}

\begin{figure}
\epsscale{1.2}
\plotone{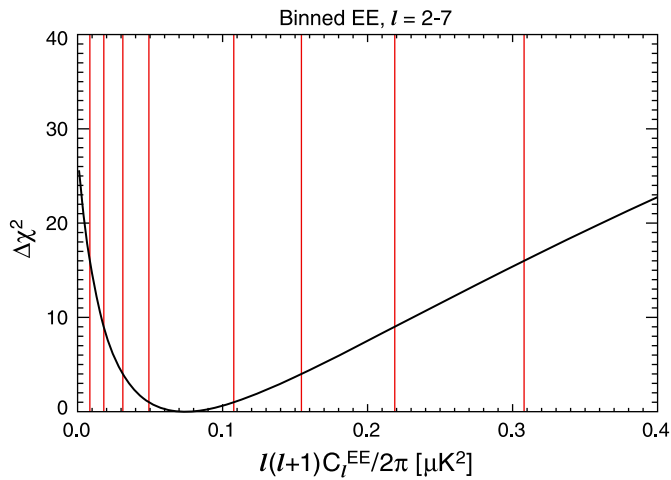}
\caption{\label{figure_low_EE_bin}
The relative $\chi^2$ for a single bin of $\ell(\ell+1)C_\ell^{\rm EE}/(2\pi)$
from $\ell$ = 2--7, conditioned on the best-fit \lcdm\ spectrum.  For simplicity,
we set $C_\ell^{\rm TE}=0$ for $\ell$ = 2--7 for this evaluation, so that the constraint 
$C_\ell^{\rm TE} \le \sqrt{C_\ell^{\rm TT} C_\ell^{\rm EE}}$ will always be satisfied.  The
vertical red lines indicate where $\Delta \chi^2$ = 1, 4, 9, and 16,
corresponding to 1, 2, 3, and 4 $\sigma$ confidence limits on EE. The 
EE=0 point in this bin
has $\Delta\chi^2 = 26.5$; additionally, setting TE=0 in the 2--7 bin
raises $\chi^2$ by 3.5
relative to the best-fit \lcdm\ TE spectrum.  Thus the full change in $\chi^2$ between
the best-fit model and the EE=TE=0 model is 30, corresponding to a 5.5$\sigma$
detection of EE power in this bin, with  $\ell(\ell+1)C_\ell^{\rm EE}/(2\pi) =
0.074^{+0.034}_{-0.025}\;\mu$K$^2$ (68\% CL).}
\end{figure}

\begin{figure*}
\epsscale{1.0}
\plotone{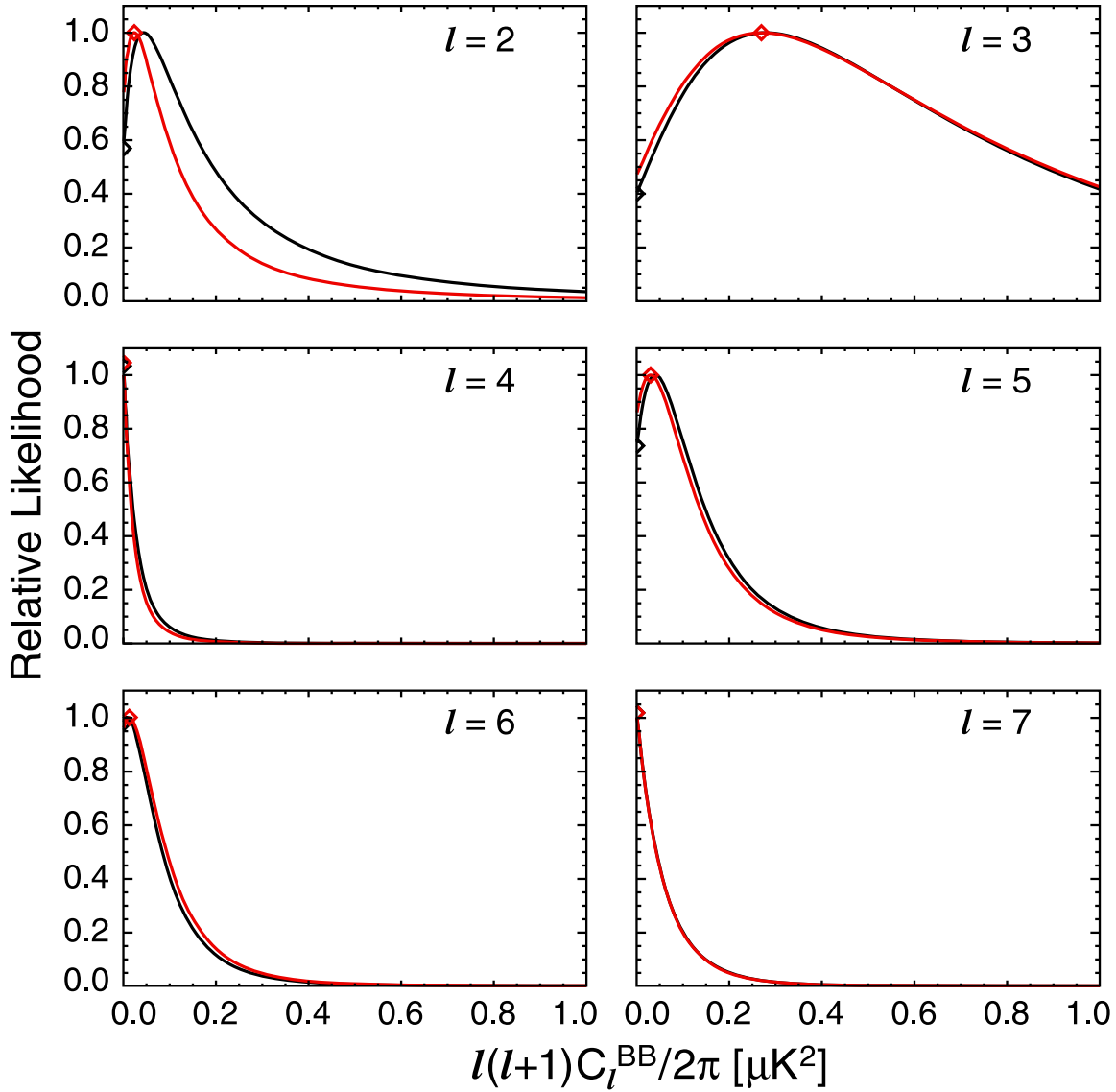}
\caption{\label{figure_low_BB}
Conditional likelihoods of $\ell(\ell+1)C_\ell^{\rm BB}/(2\pi)$ for $\ell=2$--7,
computed with the \WMAP\ likelihood code.  
The Ka, Q, and V bands contribute
to the low-$\ell$ polarized pixel likelihood.
In each panel, the black curve is the
conditional likelihood for a given multipole when all other multipoles are held
fixed at the value of the best-fit \lcdm\ model (effectively zero, except for
gravitational lensing) as indicated by the black diamonds. The red curve in each
panel is the conditional likelihood when all other multipoles are held fixed at
the maximum likelihood value of the spectrum, indicated by the red diamonds. 
The maximum likelihood spectrum was determined as stated in the caption to
\reffig{figure_low_EE}.}
\end{figure*}

\begin{figure}
\epsscale{1.2}
\plotone{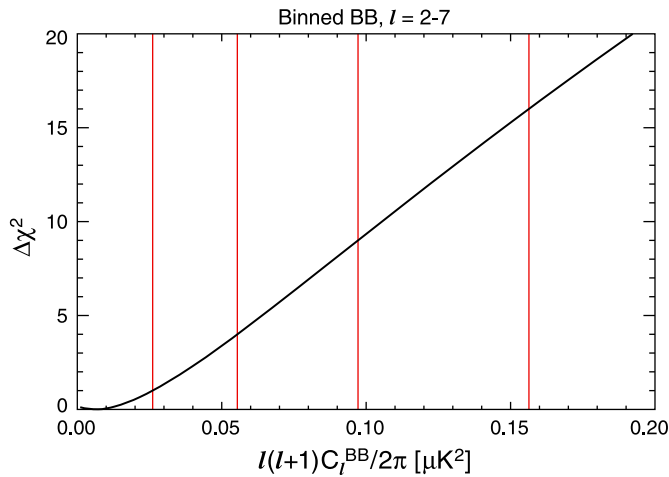}
\caption{\label{figure_low_BB_bin}
The relative $\chi^2$ for a single bin of $\ell(\ell+1)C_\ell^{\rm BB}/(2\pi)$
from $\ell = 2-7$, conditioned on the best-fit \lcdm\ spectrum.  The vertical
red lines indicate where $\Delta \chi^2$ = 1, 4, 9, and 16, corresponding to 1,
2, 3, and 4 $\sigma$ confidence limits on BB. We find an upper limit of
$\ell(\ell+1)C_\ell^{\rm BB}/(2\pi) < 0.055\;\mu$K$^2$ (95\% CL).}
\end{figure}

\begin{figure}
\epsscale{1.2}
\plotone{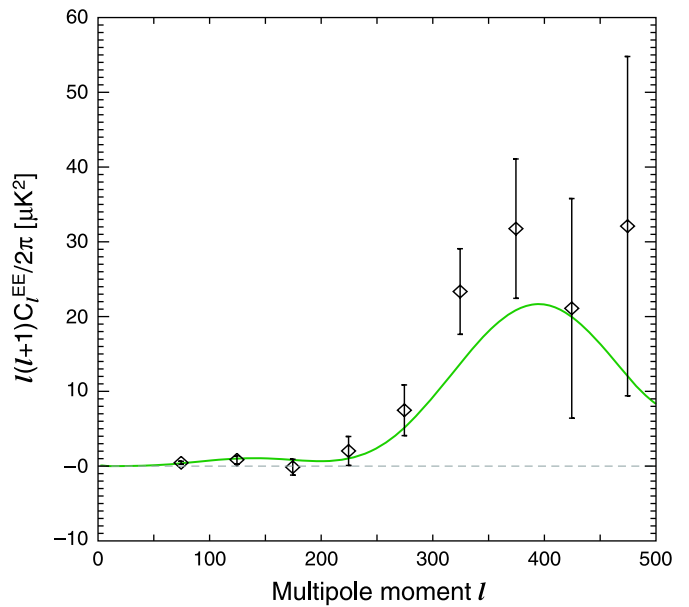}
\caption{\label{figure_high_ell_ee}
WMAP detects the high-$\ell$ EE power spectrum.  The green curve is the best fit \lcdm\
model power spectrum, and the data are a combination of Q, V, and W band data.
In the multipole range $24 \le \ell \le 800$, the detection is above $8\sigma$.
}
\end{figure}

\begin{figure}
\epsscale{1.2}
\plotone{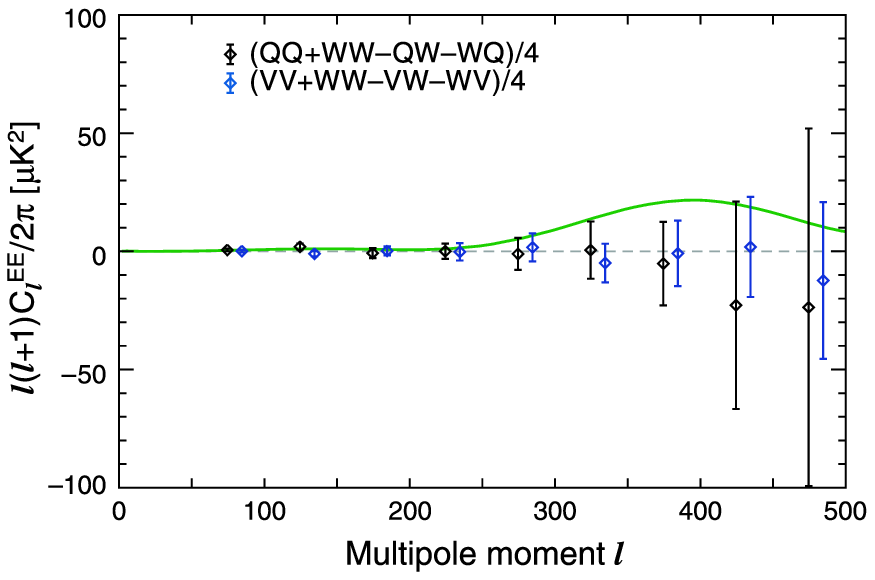}
\caption{\label{figure_high_ell_ee_diff}
The Q--W and V--W band difference spectra for EE are consistent with zero, as 
expected.   The green line is the \lcdm\ EE spectrum, which should not match
this data, but is plotted to illustrate the size of the null-test error bars, compared to the 
detected signal.
}
\end{figure}

We begin by discussing the low-$\ell$ polarization spectra, and then move on
to the high-$\ell$ EE spectrum.

The most reliable way to estimate the low-$\ell$ polarization spectra is to use
the pixel-space likelihood code to generate the posterior distributions of
individual (or binned) $C_\ell$ values.  
In the 7-year data, this code is based on a co-added Ka+Q+V map.
The most conservative, but costly,
method is to produce a Markov Chain that allows each $C_\ell$ to vary
independently; the resulting distribution of any single $C_\ell$ will be the
marginalized distribution for that multipole moment.   A Gibbs sampling
technique could also be used, but this works best with a high signal-to-noise
ratio.  However, Gibbs sampling in lower signal-to-noise  regions can be
performed successfully, as shown by \citet{jewell/etal:2009}. A much more
tractable approach is to compute the conditional likelihood in which the
likelihood of a single $C_\ell$ is evaluated while all other moments are held
fixed.  We adopt the latter approach to visualize the low-$\ell$ EE and BB
spectra.  This method has also been used in previous \WMAP\ papers 
as well as (for example) \citet{gruppuso/etal:2009}
in their verification of the 5-year \WMAP\ low-$\ell$ spectra.
In the context of parameter fitting, the estimated $C_\ell$ are
constrained to vary according to the model.

\reffig{figure_low_EE} shows the conditional likelihood for the EE multipoles
from $\ell=2$--7 for two different reference spectra. The black curves show the
likelihood of $C^{\rm EE}_\ell$ when the $C^{\rm EE}_{\ell'}$ are fixed to the
best-fit \lcdm\ model for $\ell' \ne \ell$.  The red curves are the analogous
distributions when the reference spectrum is taken to be the maximum likelihood
spectrum.
This maximum likelihood spectrum was obtained by numerical maximization
of the likelihood code for the TT, TE, EE, and BB spectra for $2\le \ell \le 10$, 
a maximization in 36 dimensions, while the spectra at $\ell > 10$ were
fixed at the best-fit \lcdm\ model.
Save for $\ell=3$ and 6, the 
likelihood curves are relatively
insensitive to the difference between these two reference spectra.  From these
curves it is clear that the majority of the statistical weight in the low-$\ell$
EE detection is at $\ell=4$, with $\ell=2$ also contributing significant power.

A standard reionization scenario would give rise to a relatively flat spectrum
in $\mathcal{C}^{\rm EE}_\ell = \ell(\ell+1)C^{\rm EE}_\ell/(2\pi)$ over the
range $\ell=2-7$, so it is of interest to evaluate the posterior distribution
of a band power with constant $\mathcal{C}$ over this range.  As shown in
Figure~\ref{figure_low_EE_bin}, we find
\be
\mathcal{C}^{\rm EE}_{2-7} = 0.074^{+0.034}_{-0.025}\;\mu{\rm K}^2 \; {\rm (68\% \; CL)}.
\ee
This result was obtained with the pixel likelihood code, and so the error bars 
include cosmic variance.
Additionally, a model with zero TE and EE power for $\ell$ = 2--7 is disfavored at
5.5$\sigma$ relative to the most-likely constant band-power in this $\ell$ range.

Figure~\ref{figure_low_BB} shows the conditional likelihood for the BB
multipoles from $\ell=2$--7 for two different reference spectra. The black
curves show the likelihood of $C^{\rm BB}_\ell$ when the $C^{\rm BB}_{\ell'}$
are fixed to the best fit \lcdm\ model (zero, except for small contributions
from lensing) for $\ell' \ne \ell$.  The red curves are the analogous
distributions when the reference spectrum is taken to be the maximum likelihood
spectrum for $\ell \le 10$ and the \lcdm\   spectrum (again, effectively zero)
for $\ell >10$.  Save for $\ell=2$, the posterior likelihood curves are
insensitive to the reference spectrum.  There is no significant detection of BB
power in any single multipole in the 7-year \WMAP\ data.  As shown in
Figure~\ref{figure_low_BB_bin}, we evaluate the posterior likelihood of a single
constant band-power from $\ell=2-7$ and find it is also consistent with zero. 
We place an upper limit of
\be
\mathcal{C}^{\rm BB}_{2-7} \le 0.055 \; \mu{\rm K}^2 \; {\rm (95\% \; CL)}
\ee
using the 7-year \WMAP\ data, which is more than a factor of 2 lower than the
5-year limit of 0.15 $\mu$K$^2$.

The high-$\ell$ EE spectrum is constructed using the polarized spectra that were
used for the TE spectrum, discussed in \refsec{sec_tp_spec}.
 \reffig{figure_high_ell_ee} plots the 7-year \WMAP\ data
on top of the \lcdm\ EE spectrum that best fits the \WMAP\ data.  Note that 
the high-$\ell$ EE spectrum is not included in the likelihood code, so the theory
curve is not a best fit to the high-$\ell$ EE spectrum.

Using error bars with cosmic variance derived from the best fit \lcdm\ model, 
we find $\chi^2 = 830.6$, for the 777 degrees of freedom in the multipole range
$24 \le \ell \le 800$.  The probability to exceed this $\chi^2$ value is 8.9\%, which is low,
but not significantly so.  For a model with no EE spectrum, $\chi^2 = 897.3$.
The difference is $\Delta \chi^2 = 66.7$, which is just over an 8 $\sigma$ detection
of the high-$\ell$ EE spectrum.

The data prefer an EE spectrum
with higher amplitude than the best-fit \lcdm\ model.  To quantify this, we find the scale factor
$\alpha$ that causes the theory EE spectrum to best fit the data, over the
multipole range $24 \le \ell \le 800$.
The scale factor is $\alpha = 1.52 \pm 0.21$.  This indicates a 2.5 $\sigma$ preference
for a higher amplitude EE spectrum, which we consider to be worth further investigation, 
but do not believe to be a significant deviation from the \lcdm\ theory.
Note that other experiments, such as QUaD \citep{brown/etal:2009},
find the EE spectrum to be consistent with the prediction of the best-fit model.

To verify that we are not seeing the power spectrum in just one band and not
the others, we take difference spectra among the Q, V, and W bands.
\reffig{figure_high_ell_ee_diff} shows two sets of difference spectra.
These spectra are consistent with zero, as expected, and demonstrate that the
EE power spectrum is present in all three frequency bands.

\subsection{\WMAP\ Likelihood Code}
\label{sec_likelihood_code}

Before discussing cosmological parameter fits in the remainder of the paper, we
review the \WMAP\ likelihood code which forms the basis for the fits.

The basic structure of the likelihood code is unchanged from the 5-year release.
For the $\ell>32$ temperature data, the model spectrum is compared to the MASTER
spectrum, described above, using a Gaussian plus log-normal approximation to the
likelihood, as described in  \citet{bond/jaffe/knox:1998} and
\citet{verde/etal:2003a}.  For $\ell \le 32$ a Blackwell-Rao estimator is used
to determine the likelihood of a model TT spectrum \citep{dunkley/etal:2009}.
This estimator encodes both the low-$\ell$ spectrum and an accurate description
of its non-Gaussian errors.  It is constructed from a set of Gibbs samples that
contain power spectra and CMB maps that are statistically consistent with the
data \citep{wandelt/larson/lakshminarayanan:2004, jewell/levin/anderson:2004,
eriksen/etal:2004e}. The 7-year input to the Gibbs chain mimics the 5-year
input: we smooth the Internal Linear Combination (ILC) map to 5$^{\circ}$
(Gaussian FWHM); degrade it to resolution $\nside=32$; and add Gaussian white
noise with 2 $\mu$K $rms$ to each pixel.  The data are masked with the KQ85y7
mask degraded to $\nside=32$, and the Gibbs sampler is run to produce input for
the Blackwell-Rao estimator; further details are given in
\citet{dunkley/etal:2009}. 

For the polarization data, we use a similar hybrid scheme: for the $\ell > 23$
TE data we compare a model spectrum to the MASTER spectrum using a Gaussian
likelihood.  (TE is the only high-$\ell$ polarization data used in the \WMAP\
likelihood code.)  For $\ell \le 23$, the likelihood of model TE, EE, and BB
spectra are obtained using a  pixel-space likelihood which is based on the
$\nside=8$ map  mentioned in \S\ref{sec_tp_spec} and described in
\citep{page/etal:2007}.

The likelihood code includes several important factors:  mode coupling due
to sky masking and non-uniform pixel weighting (due to non-uniform noise); beam
window function uncertainty, which is correlated across the entire spectrum; and
residual point source subtraction uncertainty, which is also highly correlated. 
The treatment of these effects is unchanged from the 5-year analysis
\citep{nolta/etal:2009, dunkley/etal:2009}.

Note added in revision---The results in this paper were prepared using version 4.0 of the WMAP 
likelihood.  Since the initial submission of this paper, two small errors in the likelihood code 
came to light. 1) The original computation of the TT spectrum used an incorrect monopole 
subtraction which resulted in a small amount of excess power at high $\ell$, and a 
corresponding elevation of the best-fit residual point source amplitude.  Correcting the 
monopole subtraction reduced the high-$\ell$ power slightly which produced a correspondingly 
lower residual point source amplitude, from $11.0\times10^{-3}$ to $9.0\times10^{-3} \mu$K$^2$ sr.  2) Due to a simulation 
configuration error, the TE $\fsky$ recalibration factor used in version 4.0 was 3.8\% larger than 
the final value reported in \refsec{sec_mask_fsky}.  
The first of these changes will not affect the simulations in \refsec{sec_parameter_recovery},
because they lack a monopole, and the new value for TE $\fsky$ has been used for the 
\refsec{sec_parameter_recovery} parameter recovery simulations.
The goodness of fit statistics for the TT and TE spectra
in \refsec{sec_goodness_of_fit} compare the best fit \lcdm\ theory spectrum from the version 4.0 
Markov chains (with RECFAST version 1.4.2) to the version 4.1 likelihood data.
See Appendix \ref{sec_updates} for a comparison of parameters, when estimated with the original and updated versions of the code.

\subsubsection{CAMB}

\begin{figure}
\epsscale{1.2}
\plotone{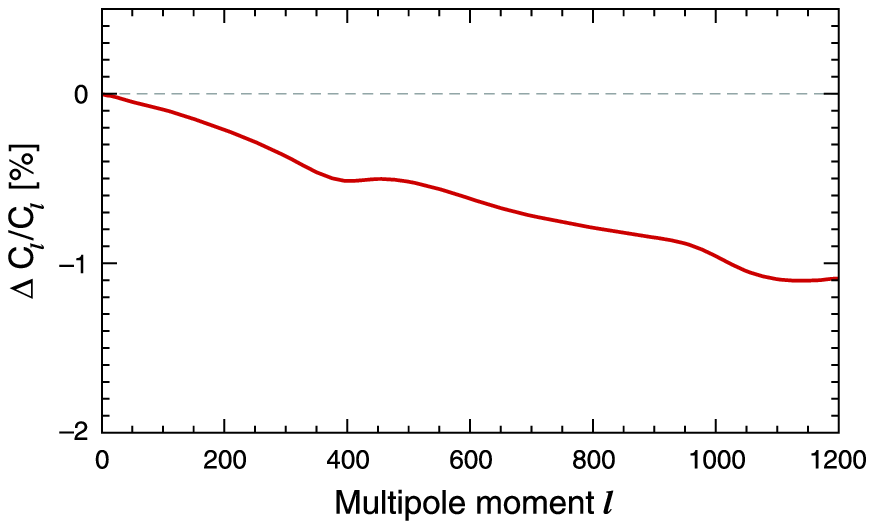}
\caption{\label{figure_recfast}
The red line in this figure represents the percentage change in the $C_\ell^{TT}$ spectrum,
due to the change in RECFAST from version 1.4.2 to version 1.5, including additional physics
of the hydrogen atom.
}
\end{figure}

For computing theoretical power spectra, we use the Code for Anisotropies in
the Microwave Background (CAMB) \citep{lewis/challinor/lasenby:2000} which is
based on the earlier code CMBFAST \citep{seljak/zaldarriaga:1996}.\footnote{We
use the November 2008 version of CAMB, which was updated to remove a bug
affecting lensed non-flat models in February 2009. The update was inadvertently
not included in our current analysis, but we have checked that the effect on
spectra is at the sub-percent level.}  Since 2008 the CAMB package has
supported improved modeling of reionization, as follows: 1) inclusion of helium
reionization, assuming that helium is singly reionized at the same time as
hydrogen and doubly reionized at $z\sim 3.5$ (this slightly lowers the redshift
of reionization for a given optical depth); 2) the width of reionization can be
varied without changing the optical depth. We use a width $\Delta z = 0.5$ as
standard. Seven-year reionization results are discussed in \S\ref{sec_delz}. 

Shortly after the 7-year WMAP data were released, a new version of CAMB was made available, incorporating 
an updated version of the code used to model recombination: RECFAST
\citep{seager/sasselov/scott:1999, seager/sasselov/scott:2000, wong/moss/scott:2008, 
scott/moss:2009}.
The parameter recovery was run with RECFAST version 1.4.2, 
and the January 2010 version of CAMB updates this to RECFAST version 1.5.
The primary change is in the optical depth, due to more accurate modeling of 
the physics of the hydrogen atom.  \reffig{figure_recfast} 
shows how much the new version of RECFAST
lowers the power spectrum, for a given set of cosmological parameters.  
The fractional lowering is largest at high $\ell$ and is about 1\%
at $\ell = 1000$.

We discuss the effect of these changes in Appendix \ref{sec_updates}.
The parameter results presented in this paper, 
and in \citet{komatsu/etal:prep}, use the original RECFAST version 1.4.2.

\section{PARAMETER RECOVERY BIAS TESTS} \label{sec_parameter_recovery}

\begin{deluxetable*}{cl}
\tablecaption{Cosmological Parameter Definitions\tablenotemark{a}\label{table_param_def}}
\tablewidth{0pt}
\tabletypesize{\footnotesize}
\tablehead{\colhead{Parameter} & \colhead{Description}}
\startdata
\multicolumn{2}{l}{Fit parameters} \\[1mm]
\ensuremath{\Omega_bh^2} &  Physical baryon density  \\
\ensuremath{\Omega_ch^2} &  Physical cold dark matter density  \\
\ensuremath{\Omega_\Lambda}   &  Dark energy density ($w=-1$ unless otherwise noted - see below) \\
\ensuremath{\Delta_{\cal R}^2}  &  Amplitude of curvature perturbations, $k_0=0.002$ Mpc$^{-1}$ \\
\ensuremath{n_s}       &  Spectral index of density perturbations, $k_0=0.002$ Mpc$^{-1}$  \\
\ensuremath{\tau}      &  Reionization optical depth  \\
\ensuremath{A_{\rm SZ}} & Amplitude of the Sunyaev-Zel'dovich spectrum\tablenotemark{b}  \\[1mm]
\multicolumn{2}{l}{Derived parameters} \\[1mm]
\ensuremath{t_0}       &  Age of the universe (Gyr) \\
\ensuremath{H_0}       &  Hubble parameter, $H_0 = 100h$ km s$^{-1}$ Mpc$^{-1}$ \\
\ensuremath{\sigma_8}   &  Amplitude of density fluctuations in linear theory, 8 $h^{-1}$ Mpc scale \\ 
\ensuremath{z_{\rm eq}}      &  Redshift of matter-radiation equality \\
\ensuremath{z_{\rm reion}}   &  Redshift of reionization \\[1mm]
\multicolumn{2}{l}{Extended parameters} \\[1mm]
\ensuremath{dn_s/d\ln{k}}   &  Running of scalar spectral index \\
\ensuremath{r}        &  Ratio of tensor to scalar perturbation amplitude, $k_0=0.002$ Mpc$^{-1}$ \\
\ensuremath{\alpha_{-1}}    &  Fraction of anti-correlated CDM isocurvature modes
(see \S\ref{sec_iso}) \\
\ensuremath{\alpha_{0}}    &  Fraction of uncorrelated CDM isocurvature modes 
(see \S\ref{sec_iso}) \\
\ensuremath{\Omega_k}   &  Spatial curvature, $\Omega_k = 1-\Omega_{\rm tot}$ \\ 
\ensuremath{w}        &  Dark energy equation of state, $w= p_{\rm DE}/\rho_{\rm DE}$ \\
\ensuremath{N_{\rm eff}}     &  Effective number of relativistic species (e.g., neutrinos) \\
\ensuremath{Y_{\rm He}}      &  Primordial Helium fraction, by mass \\
\ensuremath{\Delta_{z}}   &  Width of reionization (new parameter in CAMB, see \S\ref{sec_delz})
\enddata
\tablenotetext{a}{
Cosmological parameters discussed in this paper. A complete tabulation of
the marginalized parameter values for each of the models discussed in this
paper may be found at http://lambda.gsfc.nasa.gov.}
\tablenotetext{b}{
The Sunyaev-Zel'dovich (SZ) amplitude is not sampled in the parameter recovery simulations, 
because the SZ effect is not included in the simulation power spectrum.
\ensuremath{A_{\rm SZ}} is sampled in the 7-year WMAP Markov chains, with a
flat prior $0 < \ensuremath{A_{\rm SZ}} < 2$, but is unconstrained by the WMAP data.
See \refsec{sec_six_parameter_lcdm}.
}
\end{deluxetable*}

In this section, we describe a test for bias in the \WMAP\ parameter recovery
process. 
The parameters of the basic \lcdm\ model are: the physical baryon density,
\ensuremath{\Omega_bh^2}; the physical cold dark matter density,
\ensuremath{\Omega_ch^2}; the dark energy density, in units of the critical
density, \ensuremath{\Omega_\Lambda}; the amplitude of primordial scalar curvature
perturbations at $k = 0.002$ Mpc$^{-1}$, {\ensuremath{\Delta_{\cal R}^2}; 
the power-law spectral index of
primordial density (scalar) perturbations, \ensuremath{n_s}; and the reionization
optical depth, \ensuremath{\tau}.  The above parameters are sampled with flat
priors and are sufficiently constrained by the WMAP data that boundaries to these
priors do not have to be specified.  Nevertheless, our Markov chain code adds the 
following constraints: $0.001 < \ensuremath{\Omega_bh^2} < 0.2$, 
$0.0 < \ensuremath{\Omega_ch^2} < 0.5$, 
$0 < \ensuremath{\Omega_\Lambda}$, $0.04 < h^2 < 1.0$, and
$0.01 < \ensuremath{\tau} < 0.7$.
In this model, the Hubble constant, $H_0 =
100h$ km s$^{-1}$ Mpc$^{-1}$, is implicitly determined by the flatness
constraint, $\Omega_b + \Omega_c + \Omega_{\Lambda}=1$.  
The Sunyaev-Zel'dovich (SZ) effect
is not included in these simulations, nor is the \ensuremath{A_{\rm SZ}} parameter sampled; 
see \refsec{sec_six_parameter_lcdm} for details on this parameter.
Table~\ref{table_param_def} gives a description of the
parameters considered in this paper including both fundamental and derived
quantities.

We generate 500 simulations of \WMAP\ multi-frequency sky map data with
known cosmological parameters, then verify that we recover the correct
parameters from those data.  To our knowledge, this is the first statistical
test of the likelihood in which multiple independent realizations are combined
to test for bias at the $\sim 0.1 \sigma$ level.  Since we start the test with
simulated sky maps, this work tests: the MASTER deconvolution of the masked
pseudo-power-spectra; error propagation from maps to parameters; the code for
combining V and W band data into a single TT spectrum; the code for combining Q,
V, and W band data into a single TE spectrum; the low-$\ell$ pixel-space
likelihood codes for temperature and polarization; and the algorithm for
combining these hybrid inputs into a single likelihood per model.

There have been previous studies of the accuracy of the \WMAP\ likelihood code. 
\citet{odwyer/etal:2004} performed a Bayesian analysis of the 1-year \WMAP\ data
and found temperature power spectra largely consistent with those reported by
\citet{hinshaw/etal:2003}.  However, they pointed out that the MASTER algorithm
does not accurately represent the errors at low $\ell$.  \citet{chu/etal:2005}
investigated cosmological parameters using the statistically exact Blackwell-Rao
estimator as part of the likelihood code at low $\ell$, and found shifts of up
to $0.5\sigma$, compared to the MASTER algorithm.  These issues were addressed
by \citet{spergel/etal:2007} in the 3-year \WMAP\ analysis by using an
$N_\mathrm{side}=8$ pixel likelihood.  \citet{eriksen/etal:2007} pointed out
that the $\nside=8$ code biased $C_\ell^{TT}$ slightly high in the range $12 <
\ell < 30$, which in turn biased $n_s$ slightly low.  As a result, the final
version of \citet{spergel/etal:2003} used an $N_\mathrm{side}=16$ by code.  Since
then, the pixel likelihood code has been replaced with the Blackwell-Rao
estimator, which accurately describes the $C_\ell^{TT}$ power spectrum up to
$\ell=32$ \citep{dunkley/etal:2009}.  

The focus on $n_s$ arises because simple inflation models predict its value to
be slightly less than 1 (typically $\sim$0.96, which is termed spectral
``tilt'') and the best-fit value from previous \WMAP\ analyses is in that
range.  However, the uncertainty is such that the deviation from 1 is about
3$\sigma$, so small changes in the best-fit value can alter one's interpretation
of significance. By comparing the \WMAP\ likelihood code to a Gaussianized
Blackwell-Rao estimator, \citet{rudjord/etal:2009} report a bias in $n_s$ of
$+0.6\sigma$, which reduces the evidence for spectral tilt.  In the results
reported below, we do not find evidence for such a bias in the \WMAP\
likelihood.  In particular, using the 7-year \WMAP\ data, we find
\ensuremath{n_s = 0.963\pm 0.014} for a \lcdm\ model fit.

The pipeline for the \WMAP\ data has already been extensively tested by the
\WMAP\ team.  This testing was in progress during the planning phase of the
mission, and the pipeline continued to be refined after launch.  Power spectrum
reconstruction from maps was simulated for the first year data
\citep{hinshaw/etal:2003}.  The likelihood code was calibrated (by adjusting the
effective sky fraction, $f_{\rm sky,TT}$)  with 100,000 simulations, so that the
$\chi^2$ values reported from the likelihood could be used for goodness of fit
tests as well as model comparison \citep{hinshaw/etal:2003, verde/etal:2003}. 
Simulations of multiple years of time ordered data have shown that maps can be
reconstructed, the WMAP data calibrated from the annual dipole modulation, and
correct maps of the microwave sky recovered.

Here we present a statistical parameter extraction test: to check for bias in
our likelihood code, we construct 500 realizations of multi-year, multi-frequency sky map
data, then fit parameters from each realization independently.  The input maps
are transformed to power spectra and a likelihood code using the \WMAP\ flight
pipeline.  The simulated inputs, which are also used in
\S\ref{sec_goodness_of_fit}, are discussed in more detail in
Appendix~\ref{appendix_parameter_recovery}.

To determine if the parameter fits are biased, one could sample parameters from
a \lcdm\ Markov chain for each realization individually, then form a weighted
average.  This would be fine if the recovered parameter likelihoods were
Gaussian, but that is not guaranteed. The optimal way to combine the likelihoods
is to multiply them, and then sample from the joint distribution.  The product
of these likelihoods would represent what we know about the universe if we had
access to CMB data from 500 Hubble volumes.  Perhaps the most obvious way to
sample from this distribution is to run a Markov chain.  However, each of the
500 likelihood functions involves an independent main program, set of data
inputs, and running environment, so this solution is impractical. 

Our approach is to use importance sampling.  We want to draw samples from the
joint distribution corresponding to the sum of $N=500$ log likelihoods. 
Importance sampling draws samples from a covering distribution that is close to
the desired distribution; it then weights the samples by the ratio of
probability densities of the two distributions to correct for the difference
between the two \citep{mackay:2003}.  This approach allows us to parallelize the
processing as follows.  We generate $M=10,000$ samples from the covering
distribution, compute the model spectra for each sample and store them, then for
each of $N$ copies of the likelihood, we separately calculate the log likelihood for each 
sample spectrum.  
We add the log likelihoods, and subtract the log density of the
Gaussian at that location, to form a weight for each sample. 
This set of weighted samples is effectively the joint likelihood of
cosmological parameters over the 500 realizations.
It is not necessary to load $N$ copies of the likelihood code into memory, 
nor establish inter-process communication.

Determining a useful covering distribution is an iterative process.  We start
with a Gaussian model with no correlations between parameters, using only one
realization, and gradually add realizations (shrinking the region of interest),
updating the covariance matrix.  In the end, we use a Gaussian distribution with
a covariance matrix that is a factor of 2 larger than the covariance of the 500
combined likelihoods.  This makes the width of the distribution $\sqrt{2}$ times
``too large'' in each dimension.  Making the covariance matrix slightly larger
than the desired distribution allows us to check for the possibility of large
tails in the distribution of parameters.  Because the Gaussian sampling
distribution has tails which drop off exponentially quickly, it could (in
principle) fail to properly sample the tail of a distribution which fell less
rapidly.  To test this possibility, an array of 2-D scatter plots is made with
one cosmological parameter on each axis, and with the data points color-coded by
weight.  Visual inspection of these plots shows that the weights of the sampled
distribution are largest in the center of the sampling distribution, so that the
sampling distribution adequately covers the tails of the likelihood.  To verify
that we were not merely looking at the data in a misleading projection, we also
perform a principal component analysis on the sampled points and redisplay them
plotted on principal component axes.  The weights remain highest in the center
of the distribution.

The likelihood for these simulations differs from that used in the \WMAP\
7-year analysis in the following ways.  Most of these differences are for
computational convenience.  Because of disk space and computational time
limitations, we do not simulate 500 sets of the time ordered data and map
reconstruction; we begin with the maps.  No foregrounds are included, for
simplicity.  No beam error is included, so we do not have to simulate small
differences in deconvolution.  No unresolved point source error is included,
and we do not introduce point sources into the maps.  CAMB is run at slightly
higher accuracy, but this has a negligible effect on parameters.  The
$\nside=16$ pixel likelihood is used for $\ell \le 30$ temperature, instead of
the Gibbs likelihood, because while the Gibbs likelihood runs more rapidly, the
pixel likelihood data products are much faster to generate.  The SZ effect is
not included in the input spectrum, and so we do not attempt to fit it (unlike
the 7-year Markov chain analysis, discussed in
\refsec{sec_six_parameter_lcdm}).  While these differences mean that the
parameter recovery sims do not simulate every part of the data analysis, they
simulate a substantial portion.

Since we have 500 times more data in these parameter recovery simulations than
in the 7-year \WMAP\ data, a level of bias well below 1$\sigma$ is detectable 
in the simulations, whereas it is not in the \WMAP\ flight data.  The samples
from the importance sampling can be fit with a 6-dimensional Gaussian.  Using
this Gaussian to approximate the joint likelihood of all 500 realizations, the
input parameters have $\chi^2 = 27$ for 6 degrees of freedom, which indicates a
strong detection of a difference between the input and recovered parameters. 
The assumption of Gaussianity is reasonably good here, but the detection can
also be stated without that assumption.  Suppose one draws a line (in this
6-dimensional parameter space) between the input parameters and the mean of the
recovered parameters, and then projects all the sampled points onto that line. 
Then one can determine how far away the input parameters are from the recovered
parameters by calculating how much weight is on either side of the line from the
input parameters.  In this case, all of the weight except for one point is one one side of the
line.  Because that one point is out in the tail of the distribution, it has a weight well below average, and much less than one part in 10,000 of the weight is on the far side of
the input parameters. This indicates that the input parameters are not consistent with the
likelihood of recovered parameters. Both of these arguments indicate that the
input parameters are biased.

\begin{deluxetable*}{ccccc}
\tablecaption{Parameter Recovery Bias Test\label{param_rec}\tablenotemark{a}}
\tablewidth{0pt}
\tabletypesize{\footnotesize}
\tablehead{\colhead{Parameter\tablenotemark{b}}
& \colhead{Input value\tablenotemark{b}}
& \colhead{Measured bias\tablenotemark{c}}
& \colhead{Bias S/N\tablenotemark{d}}
& \colhead{Error bar Accuracy\tablenotemark{e}}}
\startdata
\ensuremath{10^2\Omega_bh^2} &     2.2622\phn &  \phs0.0013\phn  $\pm$ 0.0025\phn & \phs0.02   &  $-$0.035 \\
\ensuremath{\Omega_ch^2} &        0.11380    &     $-$0.00033   $\pm$ 0.00025    & $-$0.06    & \phs0.074 \\
\ensuremath{\Omega_\Lambda} &          0.72344    &  \phs0.00167     $\pm$ 0.00128    &  \phs0.06  & \phs0.039 \\
$10^9$\ensuremath{\Delta_{\cal R}^2} &   2.4588\phn &     $-$0.0160\phn$\pm$ 0.0050\phn &   $-$0.14  & \phs0.087 \\
\ensuremath{n_s} &              0.9616\phn &  \phs0.0013\phn  $\pm$ 0.0006\phn &  \phs0.09  & \phs0.020 \\
\ensuremath{\tau} &             0.08785    &     $-$0.00056   $\pm$ 0.00070    &   $-$0.04  & \phs0.042 \\ 
\enddata
\tablenotetext{a}{Parameter recovery results based on 500 Monte Carlo
simulations of 7-year \WMAP\ data fit to the 6-parameter \lcdm\ model.}
\tablenotetext{b}{Parameter and its input value in the 500 Monte Carlos realizations.}
\tablenotetext{c}{Bias measured in the composite likelihood derived from 500 MC
realizations, quoted as the mean and $rms$ of the marginalized distribution.  A
positive number indicates that the recovered value was higher than the input
value.}
\tablenotetext{d}{Measured bias divided by $rms$ of the marginalized likelihood
derived from the \WMAP\ data. The bias is less than 15\% of the 1$\sigma$ error
the 7-year data.}
\tablenotetext{e}{Fractional error in reported error bar.  We compute the standard deviation
of (output - input) / (output error), and then subtract 1.  For the 150 realizations used (this column only), we expect
fluctuations of $\pm0.058$ (1-sigma).  Our results are compatible with this.
}
\end{deluxetable*}

However, this analysis shows that the recovered parameters have very little bias compared to 
their uncertainties.
The measured level is less than 15\% of the 7-year error on each parameter, and complete 
results are given in
 \reftbl{param_rec}. Since the magnitude of the bias is small compared to the errors in the 
 7-year WMAP data, and since it will be different for different cosmological models, we do not 
attempt to remove it from the recovered parameters.

Table 2 shows that we tend to overestimate \ensuremath{n_s} by 0.09 $\sigma$. Using a different likelihood 
code on the 5-year data, \citet{rudjord/etal:2009} found a value of \ensuremath{n_s} that was 
0.6 $\sigma$ higher than our 5-year result, but with the same predicted uncertainty. Since we 
have demonstrated that our likelihood is not biased at this level, we must either conclude that a) 
there is some bias in the Rudjord et al. form of the likelihood, b) there is some undetected 
residual bias in our likelihood, or c) that both are (practically) unbiased and that different 
likelihood approximations can lead to parameter estimates that differ by this magnitude. To 
resolve this question, one should evaluate both likelihood functions on common data 
simulations, then jointly study the performance of the derived parameter ensembles.  Our 
parameter recovery simulations show that the \WMAP\ likelihood produces only a small bias 
when averaged over many CMB realizations. This does not imply that it is a better 
approximation to the exact likelihood than, e.g., the Rudjord et al. form, thus a joint comparison 
over many data realizations would be useful.  To get a very rough sense of how large a 
difference one might expect, in case c), above, we consider a toy model in which we estimate 
the variance of N random numbers from two partially overlapping subsets of the N numbers. In 
a case where 10\% of the total data sample is disjoint (i.e., 5\% in each subsample is 
independent) the two estimates of the parent variance differ, statistically, by 0.3 $\sigma$, 
where $\sigma$ is the rms of each of the subsample estimates. Thus, for two likelihood 
functions to produce parameter estimates that differ by 0.6 $\sigma$ (at 95\% confidence), the 
two functions must, in effect, be re-weighting ~5\% of the data. Given the similar construction of 
the two likelihood functions, this seems unlikely, so further study will be required to understand 
this difference.

We can also use the parameter recovery simulations to verify that our error estimates are 
correct. For each of 150 of the data realizations, we use a Markov chain to compute the 
mean and 68\% 
confidence interval for each parameter. We then examine the distribution of the quantity (output 
value - input value)/(output error) which should have a unit variance. The results are shown in 
the last column of \reftbl{param_rec}, where we find that the errors predicted by the Markov 
Chain agree with the true errors, to within the noise expected from the limited number of 
realizations.

\section{COSMOLOGICAL PARAMETERS FROM \WMAP}
\label{sec_parameters}

In this section we discuss the determination of cosmological parameters using
only the 7-year \WMAP\ data.  The measurements obtained by combining 7-year
\WMAP\ data with other cosmological data sets are presented in
\citet{komatsu/etal:prep}.  Our analysis employs the same Monte Carlo Markov
Chain (MCMC) formalism used in previous analyses \citep{spergel/etal:2003,
verde/etal:2003, spergel/etal:2007, dunkley/etal:2009, komatsu/etal:2009}.  The
MCMC formalism naturally produces parameter likelihoods that are marginalized
over all other fit parameters in the model. Throughout this paper, we quote
best-fit values as the mean of the marginalized likelihood, unless otherwise
stated (e.g., upper limits).  Lower and upper error limits correspond to the
16\% and 84\% points in the marginalized cumulative distribution, unless
otherwise stated. 

\subsection{Six-Parameter \lcdm}\label{sec_six_parameter_lcdm}

\begin{deluxetable*}{ccc}
\tablecaption{Six-Parameter \lcdm\ Fit \label{tab_lcdm}\tablenotemark{a}}
\tablewidth{0pt}
\tabletypesize{\footnotesize}
\tablehead{\colhead{Parameter} & \colhead{7-year Fit} & \colhead{5-year Fit}}
\startdata
\multicolumn{3}{l}{Fit parameters} \\[1mm]
\ensuremath{10^2\Omega_bh^2}
& \ensuremath{2.258^{+ 0.057}_{- 0.056}}
& \ensuremath{2.273\pm 0.062} \\ 
\ensuremath{\Omega_ch^2} 
& \ensuremath{0.1109\pm 0.0056}
& \ensuremath{0.1099\pm 0.0062} \\
\ensuremath{\Omega_\Lambda} 
& \ensuremath{0.734\pm 0.029}
& \ensuremath{0.742\pm 0.030} \\
\ensuremath{\Delta_{\cal R}^2} 
& \ensuremath{(2.43\pm 0.11)\times 10^{-9}}
& \ensuremath{(2.41\pm 0.11)\times 10^{-9}} \\ 
\ensuremath{n_s} 
& \ensuremath{0.963\pm 0.014}
& \ensuremath{0.963^{+ 0.014}_{- 0.015}} \\
\ensuremath{\tau} 
& \ensuremath{0.088\pm 0.015}
& \ensuremath{0.087\pm 0.017} \\[1mm]
\multicolumn{3}{l}{Derived parameters} \\[1mm]
\ensuremath{t_0} 
& \ensuremath{13.75\pm 0.13\ \mbox{Gyr}}
& \ensuremath{13.69\pm 0.13\ \mbox{Gyr}} \\
\ensuremath{H_0} 
& \ensuremath{71.0\pm 2.5\ \mbox{km/s/Mpc}}
& \ensuremath{71.9^{+ 2.6}_{- 2.7}\ \mbox{km/s/Mpc}} \\
\ensuremath{\sigma_8} 
& \ensuremath{0.801\pm 0.030}
& \ensuremath{0.796\pm 0.036} \\ 
\ensuremath{\Omega_b} 
& \ensuremath{0.0449\pm 0.0028}
& \ensuremath{0.0441\pm 0.0030} \\
\ensuremath{\Omega_c} 
& \ensuremath{0.222\pm 0.026}
& \ensuremath{0.214\pm 0.027} \\
\ensuremath{z_{\rm eq}} 
& \ensuremath{3196^{+ 134}_{- 133}}
& \ensuremath{3176^{+ 151}_{- 150}} \\
\ensuremath{z_{\rm reion}} 
& \ensuremath{10.5\pm 1.2}
& \ensuremath{11.0\pm 1.4}
\enddata
\tablenotetext{a}{Models fit to \WMAP\ data only.  See
\citet{komatsu/etal:prep} for additional constraints.}
\end{deluxetable*}

The \lcdm\ parameters used are the same as in 
\refsec{sec_parameter_recovery}, and mentioned in \reftbl{table_param_def}, 
except that \ensuremath{A_{\rm SZ}} is now also sampled.
This is a scale factor for the predicted Sunyaev-Zel'dovich spectrum 
\citep{komatsu/seljak:2002}, measured at V band,
which we add to the TT power spectrum as in \citet{spergel/etal:2007}.  
In the Markov chains, this parameter
is given a flat prior $0 < \ensuremath{A_{\rm SZ}} < 2$, but is unconstrained by the 
WMAP data, so its posterior distribution is very flat over this region.  Failing to include the SZ
effect does not significantly raise the $\chi^2$ of the fit, so only 6 parameters are needed
to provide a good fit to the WMAP power spectra, and we sample \ensuremath{A_{\rm SZ}} only
to marginalize over it.

The \lcdm\ parameters best fit to the 7-year \WMAP\ data are given in
Table~\ref{tab_lcdm}, which also lists values derived from the 5-year data for
comparison.  The results are consistent, with the 7-year measurements giving
smaller uncertainties, as expected.  The parameters that show the greatest
improvement are those that most depend on the amplitude of the third acoustic
peak and the low-$\ell$ EE polarization: \ensuremath{\Omega_bh^2},
\ensuremath{\Omega_ch^2}, and \ensuremath{\tau}, all of which are measured about
12\% more precisely.  The derived late-time matter fluctuation amplitude,
$\sigma_8$ (which depends on \ensuremath{\Omega_ch^2} and \ensuremath{\tau}), is
measured 17\% more precisely by the new data.  In \S\ref{sec_vol_change} we
consider the overall change in allowable parameter-space volume offered by the
7-year data.

As discussed in \S\ref{sec_goodness_of_fit}, this basic \lcdm\ model
continues to fit the 7-year \WMAP\ data quite well.  Indeed, none of the
additional parameters considered below provide a statistically better fit to the
7-year \WMAP\ data, after accounting for the fewer degrees of freedom in the
fits.

\subsection{Extended Cosmological Models}

In this section we examine the constraints that can be placed on augmented
\lcdm\ models (and one non-$\Lambda$ model).  In the first group we consider
parameters that introduce ``new physics'': tensor modes, a running spectral
index, isocurvature modes, spatial curvature, and non-$\Lambda$ dark energy.  In
the second group, we relax the constraints on ``standard physics'' by allowing
the effective neutrino number \& the primordial helium abundance to vary.
We also allow the reionization profile to vary.

\subsubsection{Gravitational Waves}

\begin{figure*}
\epsscale{1.0}
\plotone{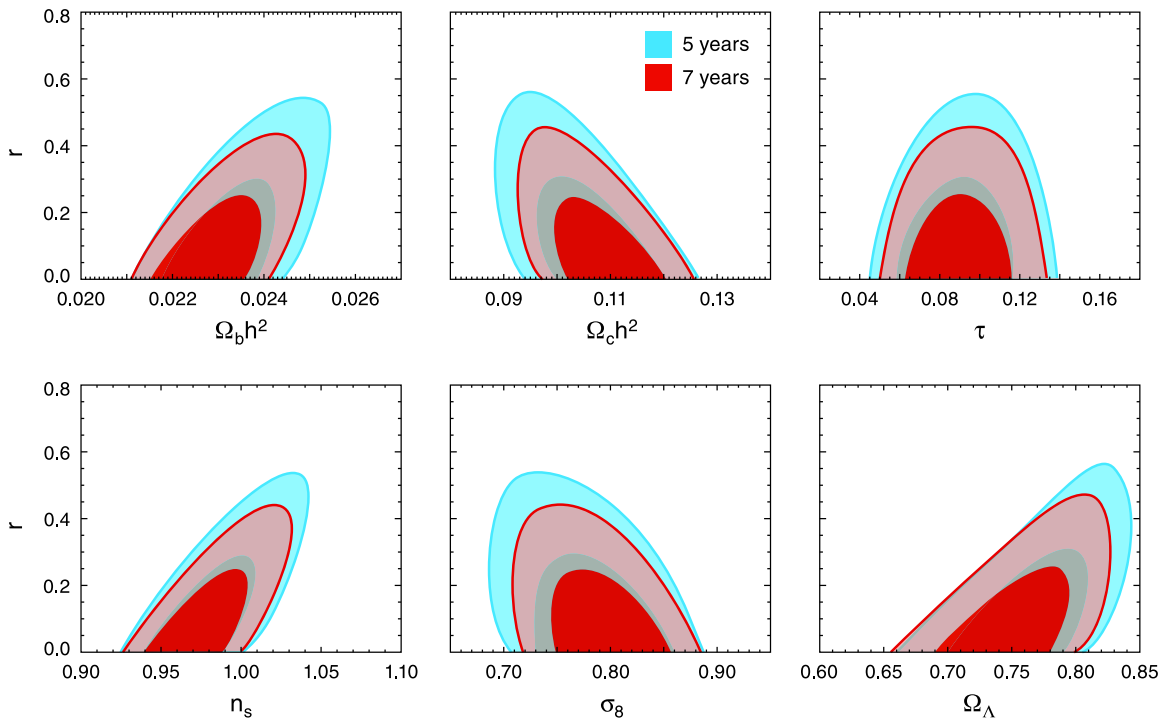}
\caption{\label{figure_2d_tensors}
Gravitational wave constraints from the 7-year \WMAP\ data, expressed in terms
of the tensor-to-scalar ratio, $r$.  The red contours show the 68\% and 95\%
confidence regions for $r$ compared to each of the 6 \lcdm\ parameters using the
7-year data.  The blue contours are the corresponding 5-year results.  We do
not detect gravitational waves with the new data; when we marginalize over the 6
\lcdm\ parameters the 7-year limit is
\ensuremath{r < 0.36\ \mbox{(95\% CL)}}, compared to the 5-year limit
of \ensuremath{r < 0.43\ \mbox{(95\% CL)}}.  Tighter limits apply
when \WMAP\ data are combined with $H_0$ and BAO constraints
\citep{komatsu/etal:prep}.}
\end{figure*}

The amplitude of tensor modes, or gravitational waves, in the early universe
may be written as
\be
\Delta_h^2(k) \equiv \frac{k^3 P_h(k)}{2\pi^2}
\ee
where $P_h(k)$ is the power spectrum of tensor perturbations at wave number $k$,
and the normalization of $P_h(k)$ is as given by \citet{komatsu/etal:2009}. This
form is comparable to the curvature perturbation amplitude,
\be
\Delta_\mathcal{R}^2(k) \equiv \frac{k^3 P_\mathcal{R}(k)}{2\pi^2}.
\ee
The dimensionless tensor-to-scalar ratio is defined as
\be
r \equiv \frac{\Delta_h^2(k)}{\Delta_\mathcal{R}^2(k)}
\ee
evaluated at $k = 0.002$ Mpc$^{-1}$.  In the Markov chain, we set a flat prior on $r$,
and require $r > 0$.

We do not detect gravitational waves from inflation with the 7-year \WMAP\ data,
however the upper limits are 16\% lower:
$\ensuremath{r < 0.36\ \mbox{(95\% CL)}}$ compared to
$\ensuremath{r < 0.43\ \mbox{(95\% CL)}}$.
Figure~\ref{figure_2d_tensors} shows the the 2-d likelihood contours for $r$ vs.
the other \lcdm\ parameters using both the 5-year and 7-year \WMAP\ data.  This
shows both the improved upper limit on $r$ and the correlations with the other
measured parameters, especially the matter densities and $n_s$.  The limits
quoted above arise from all of the power spectra measured by \WMAP\, with the
greatest power coming from the shape of the TT spectrum. 
\citet{komatsu/etal:prep} consider the constraints that arise from polarization
alone and show that the limits improve from $r<1.6$ to $r<0.93$ using the 5-year
and 7-year data, respectively.

\subsubsection{Scale Dependent Spectral Index}

\begin{deluxetable*}{cccc}
\tablecaption{Primordial Power Spectrum Constraints \label{tab_ppsc}\tablenotemark{a}}
\tablewidth{0pt}
\tabletypesize{\footnotesize}
\tablehead{\colhead{Parameter} 
& \colhead{\lcdm+Tensors}
& \colhead{\lcdm+Running}
& \colhead{\lcdm+Tensors+Running}}
\startdata
\multicolumn{4}{l}{Fit parameters} \\[1mm]
\ensuremath{\Omega_bh^2}
& \ensuremath{0.02313^{+ 0.00073}_{- 0.00072}}
& \ensuremath{0.02185^{+ 0.00082}_{- 0.00081}}
& \ensuremath{0.02221^{+ 0.00085}_{- 0.00089}} \\
\ensuremath{\Omega_ch^2}
& \ensuremath{0.1068^{+ 0.0062}_{- 0.0063}}
& \ensuremath{0.1182^{+ 0.0084}_{- 0.0085}}
& \ensuremath{0.1157^{+ 0.0086}_{- 0.0085}} \\
\ensuremath{\Omega_\Lambda}
& \ensuremath{0.757\pm 0.031}
& \ensuremath{0.688^{+ 0.052}_{- 0.051}}
& \ensuremath{0.707^{+ 0.049}_{- 0.050}} \\
\ensuremath{\Delta_{\cal R}^2}
& \ensuremath{(2.28\pm 0.15)\times 10^{-9}}
& \ensuremath{(2.42\pm 0.11)\times 10^{-9}}
& \ensuremath{(2.23^{+ 0.17}_{- 0.18})\times 10^{-9}} \\
\ensuremath{n_s}
& \ensuremath{0.982^{+ 0.020}_{- 0.019}}
& \ensuremath{1.027^{+ 0.050}_{- 0.051}}
& \ensuremath{1.076\pm 0.065} \\
\ensuremath{\tau}
& \ensuremath{0.091\pm 0.015}
& \ensuremath{0.092\pm 0.015}
& \ensuremath{0.096\pm 0.016} \\[1mm]
\ensuremath{r}
& \ensuremath{< 0.36\ \mbox{(95\% CL)}}
& \nodata
& \ensuremath{< 0.49\ \mbox{(95\% CL)}} \\
\ensuremath{dn_s/d\ln{k}}
& \nodata
& \ensuremath{-0.034\pm 0.026}
& \ensuremath{-0.048\pm 0.029} \\[1mm]
\multicolumn{4}{l}{Derived parameters} \\[1mm]
\ensuremath{t_0}
& \ensuremath{13.63\pm 0.16\ \mbox{Gyr}}
& \ensuremath{13.87^{+ 0.17}_{- 0.16}\ \mbox{Gyr}}
& \ensuremath{13.79\pm 0.18\ \mbox{Gyr}} \\
\ensuremath{H_0}
& \ensuremath{73.5\pm 3.2\ \mbox{km/s/Mpc}}
& \ensuremath{67.5\pm 3.8\ \mbox{km/s/Mpc}}
& \ensuremath{69.1^{+ 4.0}_{- 4.1}\ \mbox{km/s/Mpc}} \\
\ensuremath{\sigma_8}
& \ensuremath{0.787\pm 0.033}
& \ensuremath{0.818\pm 0.033}
& \ensuremath{0.808\pm 0.035}
\enddata
\tablenotetext{a}{Models fit to 7-year \WMAP\ data only.  See
\citet{komatsu/etal:prep} for additional constraints.}
\end{deluxetable*}

Some inflation models predict a scale dependence or ``running'' in the (nearly)
power-law spectrum of scalar perturbations.  This is conveniently parameterized
by the logarithmic derivative of the spectral index, \ensuremath{dn_s/d\ln{k}}, which
gives rise to a spectrum of the form \citep{kosowsky/turner:1995}
\be
\ensuremath{\Delta_{\cal R}^2}(k) = 
\ensuremath{\Delta_{\cal R}^2}(k_0)
\left(\frac{k}{k_0}\right)^{n_s(k_0) - 1 + \frac12 \ln(k/k_0) \ensuremath{dn_s/d\ln{k}}},
\ee
with $k_0=0.002$ Mpc$^{-1}$.  In the Markov chain, we use a flat prior on
\ensuremath{dn_s/d\ln{k}}.

We do not detect a statistically significant (i.e., $>$95\% CL) deviation from a
pure power-law spectrum with the 7-year \WMAP\ data.  The allowed range of 
\ensuremath{dn_s/d\ln{k}} is both closer to zero and has a smaller confidence range
using the 7-year data: \ensuremath{dn_s/d\ln{k} = -0.034\pm 0.026}
compared to \ensuremath{dn_s/d\ln{k} = -0.037\pm 0.028} from the
5-year data.

If we allow {\it both} tensors and running as additional primordial degrees of
freedom, the data prefer a slight negative running, but still at less than
$2\sigma$.  The joint constraint on all parameters in this model is
significantly tighter with the 7-year data (see \S\ref{sec_vol_change}).  The
7-year constraints on models with additional power spectrum degrees of freedom
are given in Table~\ref{tab_ppsc}.

\subsubsection{Isocurvature Modes}
\label{sec_iso}

\begin{deluxetable*}{cccc}
\tablecaption{Constraints on Isocurvature Modes \label{tab_iso12}\tablenotemark{a}}
\tablewidth{0pt}
\tabletypesize{\footnotesize}
\tablehead{\colhead{Parameter} 
& \colhead{\lcdm\tablenotemark{b}}
& \colhead{\lcdm+anti-correlated\tablenotemark{c}}
& \colhead{\lcdm+uncorrelated\tablenotemark{d}}}
\startdata
\multicolumn{4}{l}{Fit parameters} \\[1mm]
\ensuremath{\Omega_bh^2}
& \ensuremath{0.02258^{+ 0.00057}_{- 0.00056}}
& \ensuremath{0.02293^{+ 0.00060}_{- 0.00061}}
& \ensuremath{0.02315^{+ 0.00071}_{- 0.00072}} \\
\ensuremath{\Omega_ch^2}
& \ensuremath{0.1109\pm 0.0056}
& \ensuremath{0.1058^{+ 0.0057}_{- 0.0058}}
& \ensuremath{0.1069^{+ 0.0059}_{- 0.0060}} \\
\ensuremath{\Omega_\Lambda}
& \ensuremath{0.734\pm 0.029}
& \ensuremath{0.766\pm 0.028}
& \ensuremath{0.758\pm 0.030} \\
\ensuremath{\Delta_{\cal R}^2}
& \ensuremath{(2.43\pm 0.11)\times 10^{-9}}
& \ensuremath{(2.24\pm 0.13)\times 10^{-9}}
& \ensuremath{(2.38\pm 0.11)\times 10^{-9}} \\
\ensuremath{n_s}
& \ensuremath{0.963\pm 0.014}
& \ensuremath{0.984\pm 0.017}
& \ensuremath{0.982\pm 0.020} \\
\ensuremath{\tau}
& \ensuremath{0.088\pm 0.015}
& \ensuremath{0.088\pm 0.015}
& \ensuremath{0.089\pm 0.015} \\[1mm]
\ensuremath{\alpha_{-1}}
& \nodata
& \ensuremath{< 0.011\ \mbox{(95\% CL)}}
& \nodata \\
\ensuremath{\alpha_{0}}
& \nodata
& \nodata
& \ensuremath{< 0.13\ \mbox{(95\% CL)}} \\[1mm]
\multicolumn{4}{l}{Derived parameters} \\[1mm]
\ensuremath{t_0}
& \ensuremath{13.75\pm 0.13\ \mbox{Gyr}}
& \ensuremath{13.58\pm 0.15\ \mbox{Gyr}}
& \ensuremath{13.62\pm 0.16\ \mbox{Gyr}} \\
\ensuremath{H_0}
& \ensuremath{71.0\pm 2.5\ \mbox{km/s/Mpc}}
& \ensuremath{74.5^{+ 3.1}_{- 3.0}\ \mbox{km/s/Mpc}}
& \ensuremath{73.6\pm 3.2\ \mbox{km/s/Mpc}} \\
\ensuremath{\sigma_8}
& \ensuremath{0.801\pm 0.030}
& \ensuremath{0.784^{+ 0.033}_{- 0.032}}
& \ensuremath{0.785\pm 0.032} \\
\enddata
\tablenotetext{a}{Models fit to 7-year \WMAP\ data only.  See
\citet{komatsu/etal:prep} for additional constraints.}
\tablenotetext{b}{Repeated from Table~\ref{tab_lcdm} for comparison.}
\tablenotetext{c}{Adds curvaton-type isocurvature perturbations \citep{komatsu/etal:prep}.}
\tablenotetext{d}{Adds axion-type isocurvature perturbations \citep{komatsu/etal:prep}.}
\end{deluxetable*}

In addition to adiabatic fluctuations, where different species fluctuate in
phase to produce curvature fluctuations, it is possible to have an overdensity
in one species compensate for an underdensity in another without producing a
curvature.  These entropy, or isocurvature perturbations have a measurable
effect on the CMB by shifting the acoustic peaks in the
power spectrum.  For cold dark matter and photons, we define the field
\be
\mathcal{S}_{c,\gamma} \equiv \frac{\delta \rho_c}{\rho_c} - \frac{3 \delta \rho_\gamma}{4 \rho_\gamma}
\ee
\citep{bean/dunkley/pierpaoli:2006, komatsu/etal:2009}.  The relative amplitude
of its power spectrum is parameterized by $\alpha$,
\be
\frac{\alpha}{1 - \alpha} \equiv \frac{P_\mathcal{S}(k_0)}{P_\mathcal{R}(k_0)},
\ee
with $k_0=0.002$ Mpc$^{-1}$.

We consider two types of isocurvature modes: those which are completely
uncorrelated with the curvature modes (with amplitude \ensuremath{\alpha_{0}}),
motivated with the axion model, and
those which are anti-correlated with the the curvature modes (with amplitude
\ensuremath{\alpha_{-1}}), motivated with the curvaton model.  
For the latter, we adopt the convention in which
anticorrelation  increases the power at low multipoles
\citep{komatsu/etal:2009}.  For both \ensuremath{\alpha_{0}} and 
\ensuremath{\alpha_{-1}}, we adopt a flat prior and require $\ensuremath{\alpha_{0}} > 0$, 
$\ensuremath{\alpha_{-1}} > 0$.

The constraints on both types of isocurvature modes are given in
Table~\ref{tab_iso12}.  We do not detect a significant contribution from either
type of perturbation in the 7-year data.  The limit on uncorrelated modes
improves the most with the new data: from
\ensuremath{\alpha_{0} < 0.16\ \mbox{(95\% CL)}} to
\ensuremath{\alpha_{0} < 0.13\ \mbox{(95\% CL)}} using the 5-year and
7-year data, respectively.  Table~\ref{tab_iso12} also shows that the standard
\lcdm\ parameters are only weakly affected by the isocurvature degrees of
freedom.  \citet{komatsu/etal:prep} derive analogous constraints using a
combination of \WMAP\ plus other data.  They find limits that are roughly a
factor of two lower than the \WMAP-only limits.

\subsubsection{Spatial Curvature}

\begin{figure}
\epsscale{1.3}
\plotone{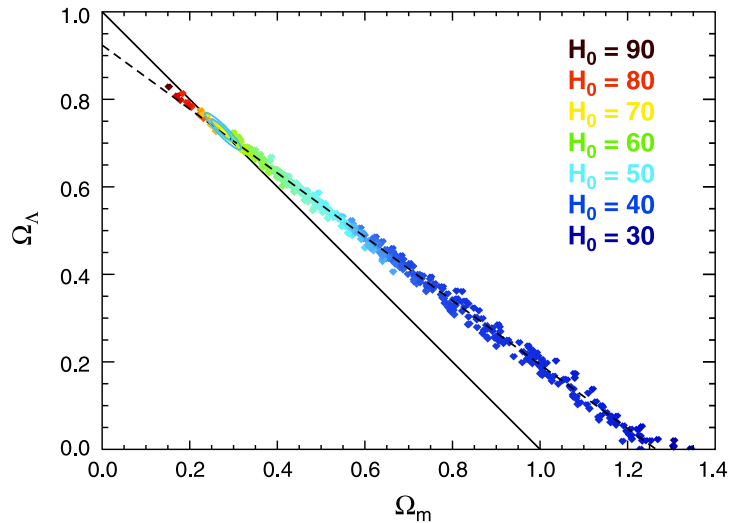}
\caption{\label{figure_oml_omm}
The range of non-flat models consistent with the \WMAP\ 7-year data.  The
plotted points are drawn from the Markov chain (\lcdm, with curvature, fit to
\WMAP\ data only); the color of each point indicates the Hubble constant for
that element in the chain, as indicated in the legend.  Due to the geometric
degeneracy, CMB data alone do not strongly constrain closed models with
$\Omega_{\Lambda}+\Omega_m>1$, provided a low Hubble constant is tolerated, see
Table~\ref{tab_geo}. The dashed line parameterizes the geometric degeneracy in
the 7-year data: $\ensuremath{\Omega_k} = -0.2654 + 0.3697\ensuremath{\Omega_\Lambda}$. 
When \WMAP\ data are combined with $H_0$ and BAO data, the result strongly
favors a flat universe:
\ensuremath{0.99<\Omega_{\rm tot}<1.01\ \mbox{(95\% CL)}}
\citep{komatsu/etal:prep}.  The joints constraints, shown as 68\% and 95\% blue
contours, provide compelling support for basic \lcdm.}
\end{figure}

\begin{deluxetable*}{cccc}
\tablecaption{Constraints on Curvature and Dark Energy \label{tab_geo}\tablenotemark{a}}
\tablewidth{0pt}
\tabletypesize{\footnotesize}
\tablehead{\colhead{Parameter} 
& \colhead{\lcdm\tablenotemark{b}}
& \colhead{O\lcdm\tablenotemark{c}}
& \colhead{$w$CDM\tablenotemark{d}}}
\tabletypesize{\footnotesize}
\startdata
\multicolumn{4}{l}{Fit parameters} \\[1mm]
\ensuremath{\Omega_bh^2}
& \ensuremath{0.02258^{+ 0.00057}_{- 0.00056}}
& \ensuremath{0.02229^{+ 0.00058}_{- 0.00057}}
& \ensuremath{0.02258^{+ 0.00063}_{- 0.00062}} \\
\ensuremath{\Omega_ch^2}
& \ensuremath{0.1109\pm 0.0056}
& \ensuremath{0.1117^{+ 0.0053}_{- 0.0055}}
& \ensuremath{0.1112\pm 0.0058} \\
\ensuremath{\Omega_\Lambda}
& \ensuremath{0.734\pm 0.029}
& \ensuremath{< 0.77\ \mbox{(95\% CL)}}
& \ensuremath{0.741^{+ 0.095}_{- 0.099}} \\
\ensuremath{\Delta_{\cal R}^2}
& \ensuremath{(2.43\pm 0.11)\times 10^{-9}}
& \ensuremath{(2.48\pm 0.11)\times 10^{-9}}
& \ensuremath{(2.43\pm 0.12)\times 10^{-9}} \\
\ensuremath{n_s}
& \ensuremath{0.963\pm 0.014}
& \ensuremath{0.955\pm 0.014}
& \ensuremath{0.964\pm 0.015} \\
\ensuremath{\tau}
& \ensuremath{0.088\pm 0.015}
& \ensuremath{0.086\pm 0.015}
& \ensuremath{0.088^{+ 0.016}_{- 0.015}} \\
\ensuremath{\Omega_k}
& \nodata
& \ensuremath{-0.080^{+ 0.071}_{- 0.093}}
& \nodata \\
\ensuremath{w}
& \nodata
& \nodata
& \ensuremath{-1.12^{+ 0.42}_{- 0.43}} \\[1mm]
\multicolumn{4}{l}{Derived parameters} \\[1mm]
\ensuremath{t_0}
& \ensuremath{13.75\pm 0.13\ \mbox{Gyr}}
& \ensuremath{15.9^{+ 2.0}_{- 1.7}\ \mbox{Gyr}}
& \ensuremath{13.75^{+ 0.29}_{- 0.27}\ \mbox{Gyr}} \\
\ensuremath{H_0}
& \ensuremath{71.0\pm 2.5\ \mbox{km/s/Mpc}}
& \ensuremath{53^{+ 13}_{- 15}\ \mbox{km/s/Mpc}}
& \ensuremath{75^{+ 15}_{- 14}\ \mbox{km/s/Mpc}} \\
\ensuremath{\sigma_8}
& \ensuremath{0.801\pm 0.030}
& \ensuremath{0.762^{+ 0.044}_{- 0.047}}
& \ensuremath{0.83\pm 0.14} \\
\enddata
\tablenotetext{a}{Models fit to 7-year \WMAP\ data only.  See
\citet{komatsu/etal:prep} for additional constraints.}
\tablenotetext{b}{Repeated from Table~\ref{tab_lcdm} for comparison.}
\tablenotetext{c}{Adds spatial curvature as a parameter, with $w\equiv-1$.  
$\ensuremath{\Omega_\Lambda}>0$ is imposed as a prior.
}
\tablenotetext{d}{Adds dark energy equation of state as a parameter, with
$\ensuremath{\Omega_k}\equiv0$.  $\ensuremath{w}>-2.5$ is imposed as a prior.}
\end{deluxetable*}

The basic \lcdm\ model of the universe is flat, with $\ensuremath{\Omega_k} =
1-\ensuremath{\Omega_{\rm tot}} = 0$.  There is a compelling theoretical case for a flat
universe in General Relativity, arising from the apparent paradox that a flat
geometry is dynamically unstable.  That is, in order for the curvature to be
acceptably small today, say $\left| \ensuremath{\Omega_k} \right| < 1$, the
curvature in the early universe had to be extraordinarily fine tuned. 
Cosmological inflation achieves this by expanding the primordial curvature
scale, if any, to super-horizon scales today.

With knowledge of the redshift of matter-radiation equality, the acoustic scale
can be accurately computed for use as a standard ruler at the epoch of
recombination.  The first acoustic peak in the CMB then provides a means to
measure the angular diameter of the acoustic scale at the surface of last
scattering.  If we have independent knowledge of the local distance-redshift
relation (the Hubble constant, $H_0$) we can infer the physical distance to the
last scattering surface, and hence the geometry of the universe.  If we assume
nothing about $H_0$ we are left with a geometric degeneracy which is illustrated
in Figure~\ref{figure_oml_omm}.

Assuming \lcdm\ dynamics, \WMAP\ data alone provide a remarkably simple
constraint on the geometry and matter-energy content in the universe. The
geometric degeneracy in the 7-year data is well described by
$\ensuremath{\Omega_k}=-0.2654+0.3697\ensuremath{\Omega_\Lambda}$ (the dashed line in
Figure~\ref{figure_oml_omm}).  We have placed a flat prior on \ensuremath{\Omega_k},
and we now also constrain $\ensuremath{\Omega_\Lambda} > 0$.
The Figure also quantitatively illustrates how
knowledge of the Hubble constant fixes the geometry, $\ensuremath{\Omega_\Lambda} +
\ensuremath{\Omega_m}$, and vice-versa.  The points in the plot are culled from
the Markov Chain that samples this model and their color is coded by the value
of the Hubble constant for that sample.  As one moves down the degeneracy line,
the Hubble constant must decrease for the model to remain consistent with the
geometry imposed by the CMB.  For a {\it flat} universe the 7-year data give
\ensuremath{H_0 = 71.0\pm 2.5\ \mbox{km/s/Mpc}} (Table~\ref{tab_lcdm}), in
excellent agreement with more traditional measurements of the Hubble constant,
e.g, \citet{riess/etal:2009}.

If we allow curvature as a parameter, the 7-year \WMAP\ data improve on the
5-year constraint by 11\%, to
\ensuremath{\Omega_k = -0.080^{+ 0.071}_{- 0.093}}.  While this result is
consistent with a flat universe, the preferred model is slightly closed and has
a relatively low Hubble constant due to the geometric degeneracy,
\ensuremath{H_0 = 53^{+ 13}_{- 15}\ \mbox{km/s/Mpc}}.  Therefore, if we impose local
distance scale measurements in the form of $H_0$ and BAO data, the limits on
curvature tighten significantly to
\ensuremath{\Omega_k = -0.0023^{+ 0.0054}_{- 0.0056}}
\citep{komatsu/etal:prep}.

\subsubsection{Non-$\Lambda$ Dark Energy}

Dark energy is believed to be driving the present-day acceleration of the
universe.  Current measurements are consistent with the dark energy being a
cosmological constant, or vacuum energy.  If it is not a cosmological constant,
then its physical density may change with the expansion of the universe.  This,
in turn, would affect the expansion history and the rate of large-scale
structure growth in the universe. The evolution of its physical density is
governed by its equation of state $w=p/\rho$ where $p$ is the pressure of the
dark energy and $\rho$ its density. The cosmological constant has an equation of
state $w=-1$.  It would be tremendously important if observations could
determine that $w \ne -1$ since that would rule out the prime candidate for the
dark energy and provide important new clues about physics.

Since the CMB primarily probes the high redshift universe ($z \sim 1000$), and
the effects of dark energy only start to dominate at relatively low redshift ($z
\sim 2$), the CMB is not especially sensitive to subtle properties of the dark
energy.  Nonetheless, meaningful constraints on the equation of state can be
inferred from the 7-year \WMAP\ data.  If we assume the universe is flat but let
$w$ be a parameter in the Friedmann equation (with a flat prior on $w$, $-2.5 < w < 0$, 
and $w'=0$),
we obtain the  constraints given in Table~\ref{tab_geo}.  In particular, the
7-year data give  \ensuremath{w = -1.12^{+ 0.42}_{- 0.43}}, which is
consistent with a cosmological constant.  \citet{komatsu/etal:prep} investigate
the constraints imposed when 7-year \WMAP\ data are combined with other
observations. With BAO and Hubble constant measurements added, they find
\ensuremath{w = -1.10\pm 0.14}, which provides compelling
limits on $w$ {\it without} using Type Ia supernovae data.  When SNe data are
included, the result becomes
\ensuremath{w = -0.980\pm 0.053}, but the quoted error
does not include systematic errors in the supernovae, which are comparable to
the statistical errors.  Accounting for this would produce a final uncertainty
that is roughly half the size of the error without SNe data.

If we relax the assumption that $w'=0$ and/or $\ensuremath{\Omega_k}=0$, the
constraints on $w$ weaken \citep{komatsu/etal:prep}.  This points to the need
for more accurate and precise measurements of the expansion history and growth
rate of structure if we are to gain further clues about dark energy from
cosmology.

\subsubsection{Neutrinos}

\begin{deluxetable*}{cccc}
\tablecaption{Constraints on Neutrino Properties \label{tab_neutrino}\tablenotemark{a}}
\tablewidth{0pt}
\tabletypesize{\footnotesize}
\tablehead{\colhead{Parameter} 
& \colhead{\lcdm\tablenotemark{b}}
& \colhead{\lcdm+\ensuremath{N_{\rm eff}}\tablenotemark{c}}
& \colhead{\lcdm+\ensuremath{\sum m_\nu}\tablenotemark{d}}}
\tabletypesize{\footnotesize}
\startdata
\multicolumn{4}{l}{Fit parameters} \\[1mm]
\ensuremath{\Omega_bh^2}
& \ensuremath{0.02258^{+ 0.00057}_{- 0.00056}}
& \ensuremath{0.02260^{+ 0.00055}_{- 0.00054}}
& \ensuremath{0.02219^{+ 0.00061}_{- 0.00062}} \\
\ensuremath{\Omega_ch^2}
& \ensuremath{0.1109\pm 0.0056}
& \ensuremath{0.162^{+ 0.041}_{- 0.038}}
& \ensuremath{0.1122\pm 0.0055} \\
\ensuremath{\Omega_\Lambda}
& \ensuremath{0.734\pm 0.029}
& \ensuremath{0.731^{+ 0.029}_{- 0.030}}
& \ensuremath{0.660^{+ 0.062}_{- 0.063}} \\
\ensuremath{\Delta_{\cal R}^2}
& \ensuremath{(2.43\pm 0.11)\times 10^{-9}}
& \ensuremath{(2.39\pm 0.11)\times 10^{-9}}
& \ensuremath{(2.50\pm 0.13)\times 10^{-9}} \\
\ensuremath{n_s}
& \ensuremath{0.963\pm 0.014}
& \ensuremath{0.992^{+ 0.022}_{- 0.023}}
& \ensuremath{0.953\pm 0.017} \\
\ensuremath{\tau}
& \ensuremath{0.088\pm 0.015}
& \ensuremath{0.088^{+ 0.014}_{- 0.015}}
& \ensuremath{0.086\pm 0.014} \\
\ensuremath{N_{\rm eff}}
& \nodata
& \ensuremath{> 2.7\ \mbox{(95\% CL)}}
& \nodata \\
\ensuremath{\sum m_\nu}
& \nodata
& \nodata
& \ensuremath{< 1.3\ \mbox{eV}\ \mbox{(95\% CL)}} \\[1mm]
\multicolumn{4}{l}{Derived parameters} \\[1mm]
\ensuremath{t_0}
& \ensuremath{13.75\pm 0.13\ \mbox{Gyr}}
& \ensuremath{11.9^{+ 1.2}_{- 1.3}\ \mbox{Gyr}}
& \ensuremath{14.09^{+ 0.25}_{- 0.26}\ \mbox{Gyr}} \\
\ensuremath{H_0}
& \ensuremath{71.0\pm 2.5\ \mbox{km/s/Mpc}}
& \ensuremath{82.6^{+ 8.9}_{- 8.7}\ \mbox{km/s/Mpc}}
& \ensuremath{65.0^{+ 4.4}_{- 4.5}\ \mbox{km/s/Mpc}} \\
\ensuremath{\sigma_8}
& \ensuremath{0.801\pm 0.030}
& \ensuremath{0.903\pm 0.077}
& \ensuremath{0.685^{+ 0.079}_{- 0.078}} \\
\enddata
\tablenotetext{a}{Models fit to 7-year \WMAP\ data only.  See
\citet{komatsu/etal:prep} for additional constraints.}
\tablenotetext{b}{Repeated from Table~\ref{tab_lcdm} for comparison.}
\tablenotetext{c}{Allows effective number of relativistic species to vary (cf.
\ensuremath{N_{\rm eff}}=3.04).  $\ensuremath{N_{\rm eff}}<10$ is imposed as a prior.}
\tablenotetext{d}{Adds neutrino mass, \ensuremath{\sum m_\nu}, as a parameter,
assuming \ensuremath{N_{\rm eff}}=3.04 and degenerate mass eigenstates.}
\end{deluxetable*}

Neutrinos affect the CMB spectrum in a variety of ways; one is by providing
relativistic degrees of freedom to the plasma prior to recombination.  Since
neutrinos, or other relativistic species, are not coupled to the photon-baryon
fluid, they free-stream out of over-densities and damp the acoustic oscillations
prior to recombination.  This action suppresses the peaks in the angular power
spectrum somewhat; the amplitude of the effect depends on the effective number
of relativistic degrees of freedom.  Using the 7-year \WMAP\ data we place a
95\% CL lower limit of \ensuremath{N_{\rm eff} > 2.7\ \mbox{(95\% CL)}} on
the number of relativistic degrees of freedom, for a flat prior
$0<\ensuremath{N_{\rm eff}}<10$.   (The standard model has $\ensuremath{N_{\rm eff}}=3.04$.) 
This new limit is 17\% higher than the 5-year limit of
\ensuremath{N_{\rm eff} > 2.3\ \mbox{(95\% CL)}} -- due to the improved
third peak measurement -- and is now quite close to the standard model value.

The mean energy of a relativistic neutrino at the epoch of recombination is
$\langle E \rangle = 0.58$ eV.  In order for the CMB power spectrum to be
sensitive to a non-zero neutrino mass, at least one species of neutrino must
have a mass in excess of this mean energy \citep{komatsu/etal:2009}.  If one
assumes that there are $\ensuremath{N_{\rm eff}}=3.04$ neutrino species with degenerate
mass eigenstates, this would suggest that the lowest total mass that could be
detected with CMB data is $\ensuremath{\sum m_\nu}\sim1.8$ eV.  Using a refined
argument, \citet{ichikawa/fukugita/kawasaki:2005} argue that one could reach
$\sim$1.5 eV.  When we add \ensuremath{\sum m_\nu} as a parameter to the \lcdm\
model (and a flat prior on the physical neutrino density, \ensuremath{\Omega_\nu h^2}, 
constrained by $\ensuremath{\Omega_\nu h^2}>0$), we obtain the fit given in 
\reftbl{tab_neutrino}, specifically
\ensuremath{\sum m_\nu < 1.3\ \mbox{eV}\ \mbox{(95\% CL)}}, which is unchanged from
the 5-year result and is slightly below the basic limits just presented. 
Note that these results come from the \WMAP\ data alone.
Tighter limits may be obtained by combining CMB data with measurements of
structure formation, as discussed in \citet{komatsu/etal:prep} and references
therein.

\subsubsection{Width of Reionization}
\label{sec_delz}

\begin{deluxetable*}{cccc}
\tablecaption{Tests of Standard Model Assumptions\label{tab_std_ass}\tablenotemark{a}}
\tablewidth{0pt}
\tabletypesize{\footnotesize}
\tablehead{\colhead{Parameter}
& \colhead{\lcdm\tablenotemark{b}}
& \colhead{\lcdm+\ensuremath{\Delta_{z}}\tablenotemark{c}}
& \colhead{\lcdm+\ensuremath{Y_{\rm He}}\tablenotemark{d}}}
\startdata
\multicolumn{4}{l}{Fit parameters} \\[1mm]
\ensuremath{\Omega_bh^2}
& \ensuremath{0.02258^{+ 0.00057}_{- 0.00056}}
& \ensuremath{0.02244\pm 0.00055}
& \ensuremath{0.02253^{+ 0.00056}_{- 0.00058}} \\
\ensuremath{\Omega_ch^2}
& \ensuremath{0.1109\pm 0.0056}
& \ensuremath{0.1117^{+ 0.0054}_{- 0.0055}}
& \ensuremath{0.1130^{+ 0.0078}_{- 0.0077}} \\
\ensuremath{\Omega_\Lambda}
& \ensuremath{0.734\pm 0.029}
& \ensuremath{0.728\pm 0.028}
& \ensuremath{0.729^{+ 0.031}_{- 0.032}} \\
\ensuremath{\Delta_{\cal R}^2}
& \ensuremath{(2.43\pm 0.11)\times 10^{-9}}
& \ensuremath{(2.46\pm 0.12)\times 10^{-9}}
& \ensuremath{(2.43\pm 0.11)\times 10^{-9}} \\
\ensuremath{n_s}
& \ensuremath{0.963\pm 0.014}
& \ensuremath{0.958^{+ 0.013}_{- 0.014}}
& \ensuremath{0.969^{+ 0.017}_{- 0.018}} \\
\ensuremath{\tau}
& \ensuremath{0.088\pm 0.015}
& \ensuremath{0.087\pm 0.015}
& \ensuremath{0.088^{+ 0.014}_{- 0.015}} \\[1mm]
\ensuremath{Y_{\rm He}}
& \nodata
& \nodata
& \ensuremath{0.28^{+ 0.14}_{- 0.15}} \\[1mm]
\multicolumn{4}{l}{Derived parameters} \\[1mm]
\ensuremath{t_0}
& \ensuremath{13.75\pm 0.13\ \mbox{Gyr}}
& \ensuremath{13.77^{+ 0.13}_{- 0.12}\ \mbox{Gyr}}
& \ensuremath{13.69\pm 0.16\ \mbox{Gyr}} \\
\ensuremath{H_0}
& \ensuremath{71.0\pm 2.5\ \mbox{km/s/Mpc}}
& \ensuremath{70.5\pm 2.4\ \mbox{km/s/Mpc}}
& \ensuremath{70.9\pm 2.5\ \mbox{km/s/Mpc}} \\
\ensuremath{\sigma_8}
& \ensuremath{0.801\pm 0.030}
& \ensuremath{0.802\pm 0.030}
& \ensuremath{0.820^{+ 0.053}_{- 0.054}}
\enddata
\tablenotetext{a}{Models fit to 7-year \WMAP\ data only.  See
\citet{komatsu/etal:prep} for additional constraints.}
\tablenotetext{b}{Repeated from Table~\ref{tab_lcdm} for comparison.}
\tablenotetext{c}{Allows width of reionization to vary.  The constraints on 
\ensuremath{\Delta_{z}} are limited by the prior $0.5 < \ensuremath{\Delta_{z}} < 15$
and so are not listed in the table.
}
\tablenotetext{d}{Allows primordial helium mass fraction to vary (cf.
\ensuremath{Y_{\rm He}}=0.24).}
\end{deluxetable*}

Effective with the March 2008 version of the code
CAMB \citep{lewis:2008},  a new
parameter has been added which allows users to vary the reionization profile
while holding the total optical depth fixed.  The basic profile is a smooth ramp
in redshift space and the parameter, \ensuremath{\Delta_{z}}, changes the slope of
the ramp about its midpoint in such a way as to preserve total optical depth.

We have added \ensuremath{\Delta_{z}} as a parameter to the basic \lcdm\ model,
with a flat prior in the range $0.5<\ensuremath{\Delta_{z}}<15$, and
present the results in \reftbl{tab_std_ass}.

\subsubsection{Primordial Helium Abundance}

Helium is thought to be synthesized in the early universe via Big Bang
Nucleosynthesis (BBN).  Given the \WMAP\ measurement of the baryon-to-photon
ratio, $\eta$, the BBN-predicted yield for helium is \ensuremath{Y_{\rm He}}=0.249
\citep{steigman:2007}.  To date, the best technique for measuring the primordial
abundance has been to observe stars in HII regions: in these systems, the helium
abundance as a function of metallicity can be observed and the relation can be
regressed to zero metallicity, which is presumed to give the primordial
abundance \citep{gruenwald/steigman/viegas:2002, izotov/thuan:2004,
olive/skillman:2004, fukugita/kawasaki:2006, peimbert/luridiana/peimbert:2007}
as reviewed by \citet{steigman:2007}.

Primordial helium affects the time profile of recombination which, in turn,
affects the CMB angular power spectrum, especially the third acoustic peak.  With
\WMAP's improved measurement of this peak, it is now possible to let
\ensuremath{Y_{\rm He}} be a fitted parameter in the \lcdm\ model.  We use a flat prior
with $0.01 < \ensuremath{Y_{\rm He}} < 0.8$.  We present the
results of this fit in Table~\ref{tab_std_ass} and call out the helium abundance
specifically: \ensuremath{Y_{\rm He} = 0.28^{+ 0.14}_{- 0.15}}.  This result
is consistent with the BBN prediction and suggests the existence of pre-stellar
helium, at the $\sim 2 \sigma$ level.  \citet{komatsu/etal:prep} consider the
constraints that can be applied when higher-resolution CMB data are included in
the fit.  They find that the combined CMB data produce, for the first time,
evidence for pre-stellar helium at $>3\sigma$.

\subsection{Volume Change}
\label{sec_vol_change}

\begin{deluxetable*}{lcc}
\tablecaption{Parameter-Space Volume Reduction\label{tab_vol}\tablenotemark{a}}
\tablewidth{0pt}
\tabletypesize{\footnotesize}
\tablehead{\colhead{Model}
& \colhead{Dimension}
& \colhead{Ratio\tablenotemark{b}}}
\startdata
\lcdm\ (Table~\ref{tab_lcdm})                             & 6 & 1.5 \\
\lcdm+tensors   (Table~\ref{tab_ppsc})                    & 7 & 1.9 \\
\lcdm+running (Table~\ref{tab_ppsc})                      & 7 & 1.7 \\
\lcdm+tensors+running (Table~\ref{tab_ppsc})              & 8 & 3.0 \\
\lcdm+anticorrelated isocurvature (Table~\ref{tab_iso12}) & 7 & 1.9 \\
\lcdm+uncorrelated isocurvature (Table~\ref{tab_iso12})   & 7 & 1.9 \\
\lcdm+massive neutrinos (Table~\ref{tab_lcdm})            & 7 & 1.8 \\
o\lcdm\ (Table~\ref{tab_geo})                             & 7 & 1.8 \\
$w$CDM (Table~\ref{tab_geo})                              & 7 & 1.5
\enddata
\tablenotetext{a}{The relative change in allowable parameter-space volume when
models are fit to the 7-year \WMAP\ data in place of the 5-year data.  For a
given model, the allowable volume is defined as the square root of the
determinant of the parameter covariance matrix, as obtained from the Markov
chains.  The basic set of 6 parameters compared in the first row are
$\{\ensuremath{\Omega_bh^2}, \ensuremath{\Omega_ch^2},  \ensuremath{\Omega_\Lambda},
\ensuremath{\Delta_{\cal R}^2}, \ensuremath{n_s}, \ensuremath{\tau}\}$. Additional
parameters are as noted in the first column.}
\tablenotetext{b}{The ratio of allowable parameter-space volume: 5-year over
7-year, when fit to \WMAP\ data only.}
\end{deluxetable*}

The tables presented in this section quote parameter uncertainties for
marginalized 1-dimensional likelihood profiles.  With the 7-year and 5-year
results side by side, one could infer the improvement in precision for any given
parameter in any given model fit, and we have called out examples in the text. 
But it is difficult to measure the overall improvement in the model fits from
this presentation.  A better measure is given by comparing the allowable volume
in $N$-dimensional parameter space for each of the models.

As a proxy for allowable volume, we compute the square root of the determinant
of the parameter covariance matrix for each model, using data from the Markov
chains.  For example, with the 6-parameter \lcdm\ model, we compute the
$6\times6$ covariance matrix directly from the chain samples.  The allowable
volume ratio is defined as the ratio of the square root of the 5-year
determinant to the corresponding 7-year value.  Models with more fit parameters
have more rows and columns in their covariance matrix.  While this proxy is only
proportional to the volume if the parameter distributions are Gaussian, the error in this
approximation will be similar for both the 5-year and 7-year data sets, so a
comparison is still valid.

Table~\ref{tab_vol} gives the change in allowable parameter-space volume as a
ratio of the 5-year to 7-year value.  These results are based on fits to \WMAP\
data only.  Overall, the 6-parameter \lcdm\ model is measured a factor of 1.5 more
precisely with the 7-year data while the model with 2 additional parameters,
tensors plus a running spectral index, is measured a factor of 3 times more
precisely.  Models with one additional parameter typically improve by factors of 1.8 to 1.9.

\section{GOODNESS OF FIT}
\label{sec_goodness_of_fit}

Given a best-fit model from the MCMC analysis, we can ask how well the model fits
the data.  Given that the likelihood function is non-Gaussian, answering the
question is not as straightforward as testing the $\chi^2$ per degree of freedom of
the best-fit model.  Instead we resort to Monte Carlo simulations and compare the
absolute likelihood obtained from fitting the flight data to an ensemble of
simulated values.

For testing goodness of fit, we generate 500 realizations of the 7-year sky map
data that include the CMB signal and instrument noise.  These are the same realizations
as were used for parameter recovery in \refsec{sec_parameter_recovery}, and are
discussed in more detail in Appendix \ref{appendix_parameter_recovery}.

For each realization of a 7-year data set, we constructed the likelihood function
appropriate to those data.  This required forming the high-$\ell$ MASTER spectra
and the low-resolution sky maps used in the code.  (For this study we did not
employ Gibbs sampling for the low-$\ell$ TT likelihood, rather we used a direct
pixel-space code that was computationally slower than the Blackwell-Rao estimate
per likelihood evaluation, but it required less setup overhead per data
realization.)  For goodness of fit testing, we evaluated the likelihood of the
input \lcdm\ model for each data realization.  

Due to its hybrid nature, the likelihood produces several components that need
to be combined to obtain the full likelihood.  The components of most interest
to goodness of fit testing are the high-$\ell$ TT and TE portions, which cover the
bulk of the multipole range and are the most straightforward to interpret. 
Recall the the high-$\ell$ TT component contains both a Gaussian and a log-normal
contribution, as per equation~11 of \citet{verde/etal:2003}
\be
\ln\mathcal{L} = \frac{1}{3} \ln\mathcal{L}_{\rm Gauss}
               + \frac{2}{3} \ln\mathcal{L}'_{\rm LN}.
\ee
Here the first term can be compared to $\chi^2$, as per equation~6 of
\citet{verde/etal:2003}
\be
\ln\mathcal{L}_{\rm Gauss} \propto -\frac{1}{2} \sum_{\ell\ell'} 
 (\hat{\mathcal{C}}_\ell - \mathcal{C}^{\rm th}_\ell) 
 Q_{\ell\ell'}
 (\hat{\mathcal{C}}_{\ell'}-\mathcal{C}^{\rm th}_{\ell'}).
\ee
where $Q$ is the inverse covariance matrix of the observed power spectrum
$\hat{\mathcal{C}}$, and $\mathcal{C}^{\rm th}$ is the model spectrum.  The
high-$\ell$ portion of the TE likelihood includes only the Gaussian component,
which is a good approximation given the lower signal-to-noise ratio of the TE
data.  In the following, we report on the distribution of $-2\ln\mathcal{L}$
which we call the effective $\chi^2$.

We compare the distribution of effective $\chi^2$ values for the high-$\ell$ 
TT portion of the likelihood, which contains
1170 multipoles from $\ell=31$--1200, to a 
$\chi^2$ distribution with 1170 degrees of freedom.
The agreement between the two distributions is good.  We
tentatively attribute a small shape difference between them to the
non-Gaussian component in the likelihood.  The effective $\chi^2$ for the 7-year
flight TT spectrum is 1227 for 1170 degrees of freedom, after marginalizing over point
sources and the SZ spectrum (which are not in the simulations).  
According to the Monte Carlo distribution, $48/500$ of the
realizations had a higher effective $\chi^2$, indicating that the flight data
are reasonably well fit by the \lcdm\ model spectrum.

We perform a similar comparison for
the high-$\ell$ TE data, which covers the multipole range $\ell=24$--800.  
One point of note is that we have adjusted our
empirical calibration of $f_{\rm sky,TE}$ as a result of these simulations: the
new value is 1.011 times larger than we used in the 5-year analysis, which is
equivalent to over-estimating the 5-year TE errors by 1.1\%.  The re-calibrated
Monte Carlo distribution tracks the pure $\chi^2$ distribution, consistent with
the high-$\ell$ TE likelihood being Gaussian.  The effective $\chi^2$ for the
7-year flight TE spectrum is 807 for 777 degrees of freedom (again, after marginalization
over point sources and the SZ spectrum).
According to the Monte Carlo distribution, $113/500$ of the simulations had a higher effective 
$\chi^2$, which could easily happen by random chance.  This indicates that the \lcdm\ theory
yields a TE spectrum that fits the data well.

\section{CONCLUSIONS}
\label{sec_conclusion}

We present the angular power spectra derived from the 7-year \WMAP\ sky maps and
discuss the cosmological conclusions that can be inferred from \WMAP\ data
alone.  

With the 7-year data, the temperature (TT) spectrum measurement is now limited
by cosmic variance for multipoles $\ell<548$, and the signal-to-noise ratio per
multipole exceeds unity for $\ell<919$.  In a band-power of width $\Delta l=10$
the signal-to-noise ratio exceeds unity to $l=1060$.  The third acoustic peak in
the TT spectrum is now well measured by \WMAP.  In the context of a flat \lcdm\
model, this improvement allows us to place tighter constraints on the matter
density from \WMAP\ data alone,
\ensuremath{\Omega_mh^2 = 0.1334^{+ 0.0056}_{- 0.0055}}, and on the epoch of
matter-radiation equality, \ensuremath{z_{\rm eq} = 3196^{+ 134}_{- 133}}. The
temperature-polarization (TE) spectrum is detected in the 7-year data with a
significance of $20\sigma$, compared to $13\sigma$ with the 5-year data.  We now
detect the second dip in the TE spectrum near $\ell\sim450$ with high
confidence.  The TB and EB spectra remain consistent with zero in the 7-year
data.  This demonstrates low systematic errors in the data and is used to place
33\% tighter limits on the rotation of linear polarization due to
parity-violating effects: $\Delta \alpha=-1.1^\circ \pm 1.4^\circ
\mathrm{(stat.)} \pm 1.5^\circ \mathrm{(sys.)}$ \citep{komatsu/etal:prep}.  The
low-$\ell$ EE spectrum, a measure of the optical depth due to reionization, is
detected at $5.5\sigma$ significance when averaged over $\ell=2$--7:
$\ell(\ell+1)C_\ell^{EE}/(2\pi) = 0.074^{+0.034}_{-0.025}$ $\mu$K$^2$ (68\% CL).
The high-$\ell$ EE spectrum in the range $24\le \ell \le 800$ is detected at over 8 $\sigma$.
The BB spectrum, an important probe of gravitational waves from inflation,
remains consistent with zero; when averaged over $\ell=2$--7,
$\ell(\ell+1)C_\ell^{BB}/(2\pi) < 0.055$ $\mu$K$^2$ (95\% CL). The upper limit
on tensor modes from polarization data alone is a factor of 2 lower with the
7-year data than it was using the 5-year data \citep{komatsu/etal:prep}.

The data remain consistent with the simple \lcdm\ model.  The best-fit \lcdm\
parameter values are given in Table~\ref{tab_lcdm}; the TT spectrum from this
fit has an effective $\chi^2$ of 1227 for 1170 degrees of freedom, with a
probability to exceed of 9.6\%. The allowable volume in the 6-dimensional space
of \lcdm\ parameters has been reduced by a factor of 1.5 relative to the 5-year
volume.  Most models with one additional parameter beyond \lcdm\ see volume
reduction factors of 1.8--1.9, while the \lcdm\ model that
allows for tensor modes and a running scalar spectral index has a factor of 3
lower volume when fit to the 7-year data.  We test the parameter recovery
process for bias and find that the scalar spectral index, \ensuremath{n_s}, is
biased high, but only by $0.09 \sigma$, while the remaining parameters are
biased by $<0.15 \sigma$.

The improvement in the third peak measurement leads to tighter lower limits from
\WMAP\ on the number of relativistic degrees of freedom (e.g., neutrinos) in the
early universe: \ensuremath{N_{\rm eff} > 2.7\ \mbox{(95\% CL)}}.  Also,
using \WMAP\ data alone, the primordial helium mass fraction is found to be
\ensuremath{Y_{\rm He} = 0.28^{+ 0.14}_{- 0.15}}, and with data from
higher-resolution CMB experiments included, \citet{komatsu/etal:prep} establish
the existence of pre-stellar helium at $>3\sigma$.

\acknowledgements

The \WMAP\ mission is made possible by the support of the NASA Science Mission
Directorate. This research has made use of NASA's Astrophysics Data System Bibliographic
Services. 
Some of the results in this
paper have been derived using the HEALPix \citep{gorski/etal:2005} package.  We
acknowledge use of the CAMB \citep{lewis/challinor/lasenby:2000}
and CMBFAST \citep{seljak/zaldarriaga:1996} packages.

\appendix

\section{PARAMETER RECOVERY SIMULATIONS}
\label{appendix_parameter_recovery}

This appendix describes the configuration of the set of 500 simulations that was used
for checking parameter recovery and the $\chi^2$ values from the likelihood.

The mask is the KQ85y7 mask, which lets through 78.3\% of the sky.

For instrument noise, we employ a 2-step process in which we generate
uncorrelated noise at high resolution and combine it with low resolution
correlated noise.  The noise was constructed from the 7-year $\nside=512$ and
1024 $N_\mathrm{obs}$ maps, the 7-year $\nside=16$ covariance matrices, the
7-year $\sigma_0$ values, and the 7-year synchrotron cleaning factors, given in
\citet{gold/etal:prep}.  For each year and nine DAs (Ka1--W4), a high
resolution uncorrelated noise map was made with I, Q, U components at
$\nside=512$ and and 1024, including the QU correlations within each pixel.
The noise maps are generated on a single-year, single-DA basis so we can mimic
the construction of the flight spectra and likelihood function.  A correlated
noise map was made at $\nside=16$.  The $\sigma_0$ values for the polarization
portion of these maps were increased by a factor of $1/(1-a_1)$, where $a_1$ is
the fraction of the K band map which has been removed to avoid synchrotron
contamination \citep{gold/etal:prep}.  This accounts for the increased noise
due to the synchrotron template subtraction.  We combined the low resolution
($\nside=16$) and high resolution ($\nside = 512$ or 1024) maps by subtracting
off the mean of the high resolution noise within each low resolution pixel, and
then added the low resolution noise to all high resolution pixels within that
low resolution pixel.  This process provides a high resolution noise
realization that has the proper low resolution noise correlations when it is
binned.  Note that the $\nside=512$ and 1024 maps have different noise
realizations, but this will have no effect on the resultant likelihoods,
because the $\nside=512$ maps are used for polarization and the $\nside=1024$
maps are used for temperature, following the procedure used in the standard
\WMAP\ pipeline.  

The CMB signal is assumed to be Gaussian, and random-phase, and so its
statistical properties are completely defined by a power spectrum.  The
parameter recovery simulations all use the same power spectrum, which was
derived from the best fit to a 5-year \lcdm\ Markov chain, with \WMAP, Baryon
Acoustic Oscillation, and Supernova data.  The parameters used are $\Omega_b
h^2 = 0.0022622$, $\Omega_c h^2=0.1138$, $H_0 = 70.234 \; \mathrm{km}\,
\mathrm{s}^{-1} \mathrm{Mpc}^{-1}$, $\Delta_\mathcal{R}^2 =
2.4588\times10^{-9}$, $n_s = 0.9616$, $\tau=0.08785$.  The gravitational
lensing signal is treated as Gaussian, and the effects on the temperature and
polarization power spectra are included.  However, the BB spectrum has been
zeroed, for consistency with a map-making simulation done previously.  Zeroing
the BB spectrum will have no effect on parameter recovery, since our simple
\lcdm\ model has a tensor to scalar ratio of 0, and therefore an undetectable
BB spectrum.  Each CMB realization has the same theoretical power spectrum, but
different cosmic variance.  The different DAs in a given realization all see
the same CMB sky, but with different smoothing, and different noise.  The
smoothing used for each DA is a circular beam response based on the appropriate
7-year beam transfer function \citep{jarosik/etal:prep}.  

These simulations do not include foregrounds, the Sunyaev-Zel'dovich effect, 
unresolved point sources, or beam uncertainty.

The analysis of the parameter recovery simulations uses the same parameters as
went into the maps, in the case of $N_\mathrm{obs}$, beam profiles, and usage
of gravitational lensing.  However, for the $\sigma_0$ and synchrotron cleaning
coefficients, the simulations were produced with the 7-year values, and
analyzed with the previous 5-year values.  Note that this is a small
difference, because the 7-year values are not very different from the 5-year
values.

The parameter recovery importance sampling used the September 2008 version of
CAMB, and so does not include the bug fix for the proton mass error.  This has
no effect on the simple \lcdm\ model explored here.

\section{RECFAST AND WMAP LIKELIHOOD UPDATES}
\label{sec_updates}

\begin{deluxetable*}{ccc}
\tablecaption{Likelihood Updates\label{tab_like_update}\tablenotemark{a}}
\tablewidth{0pt}
\tabletypesize{\footnotesize}
\tablehead{\colhead{Parameter}
& \colhead{Updated\tablenotemark{b}}
& \colhead{Original\tablenotemark{c}}}
\startdata
\multicolumn{3}{l}{Fit parameters} \\[1mm]
\ensuremath{10^2\Omega_bh^2} & $ 2.249_{-0.057}^{+0.056} $ & \ensuremath{2.258^{+ 0.057}_{- 0.056}} \\
\ensuremath{\Omega_ch^2} & $ 0.1120\pm 0.0056 $ & \ensuremath{0.1109\pm 0.0056} \\
\ensuremath{\Omega_\Lambda} & $ 0.727_{-0.029}^{+0.030} $ & \ensuremath{0.734\pm 0.029} \\
\ensuremath{\Delta_{\cal R}^2} & $ (2.43\pm 0.11)\times 10^{-9} $ & \ensuremath{(2.43\pm 0.11)\times 10^{-9}} \\
\ensuremath{n_s} & $ 0.967\pm 0.014 $ & \ensuremath{0.963\pm 0.014} \\
\ensuremath{\tau} & $ 0.088\pm 0.015 $ & \ensuremath{0.088\pm 0.015} \\[1mm]
\multicolumn{3}{l}{Derived parameters} \\[1mm]
\ensuremath{t_0} & $ 13.77\pm 0.13 \;{\rm Gyr}$ & \ensuremath{13.75\pm 0.13\ \mbox{Gyr}} \\
\ensuremath{H_0} & $ 70.4\pm 2.5 \;{\rm km/s/Mpc}$ & \ensuremath{71.0\pm 2.5\ \mbox{km/s/Mpc}} \\
\ensuremath{\sigma_8} & $ 0.811_{-0.031}^{+0.030} $ & \ensuremath{0.801\pm 0.030} \\
\ensuremath{\Omega_b} & $ 0.0455\pm 0.0028 $ & \ensuremath{0.0449\pm 0.0028} \\
\ensuremath{\Omega_c} & $ 0.228\pm 0.027 $ & \ensuremath{0.222\pm 0.026} \\
\ensuremath{z_{\rm reion}} & $ 10.6\pm 1.2 $ & \ensuremath{10.5\pm 1.2} 
\enddata
\tablenotetext{a}{7-year \lcdm\ parameters, to illustrate the differences between two versions of RECFAST and two versions of the likelihood code.  The 6-parameter model and priors are the same as previously used in Sections \ref{sec_parameter_recovery} and \ref{sec_parameters}, and mentioned in \reftbl{tab_lcdm}.  Here, we also marginalize over the SZ effect, as in \refsec{sec_six_parameter_lcdm}.}
\tablenotetext{b}{The updated version of the parameters, based on RECFAST 1.5 and version 4.1 of the \WMAP\ likelihood.  This version is  more accurate, but was not available when the chains in the rest of this paper were run.}
\tablenotetext{c}{The original version of the parameters, based on RECFAST 1.4.2 and version 4.0 of the \WMAP\ likelihood.  This is the code configuration used for the chains reported in this paper.}
\end{deluxetable*}

As mentioned in \refsec{sec_likelihood_code}, since the original version of this paper, 
there have been two updates to the 
parameter estimation code: an improvement in RECFAST from version 1.4.2 to version 1.5, which includes an improved model of the hydrogen atom; and a small bug fix in the \WMAP\ likelihood code, changing its version from 4.0 to 4.1.

We have rerun the \lcdm\ parameter fits 
using the version 4.1 likelihood, and find parameter changes of order 0.1 $\sigma$.
The effect on parameters of updating to the new version of RECFAST in CAMB is also of order 0.1 $\sigma$.   The largest combined changes are 
an increase in \ensuremath{\sigma_8} of 0.35 $\sigma$, and
an increase in the spectral index \ensuremath{n_s} of 0.26 $\sigma$;
\reftbl{tab_like_update} provides more detailed information on several original and updated 
parameter values.

\bibliographystyle{wmap}
\bibliography{wmap}

\end{document}